\documentclass[%
reprint,
aps,nofootinbib,
notitlepage,
superscriptaddress,
twocolumn,
nobibnotes,
 amsmath,
 amssymb,
 prl,
]{revtex4-2}
\usepackage{graphicx}
\usepackage{dcolumn}
\usepackage{bm}
\usepackage{braket}
\usepackage{float}
\usepackage{lipsum}
\usepackage{graphicx}
\usepackage{dcolumn}
\usepackage{bm}
\usepackage{natbib} 
\usepackage{hyperref} 
\usepackage[caption=false]{subfig}
\usepackage{bm}

\hypersetup{
	colorlinks = true,
    linkcolor = Maroon,
    urlcolor  = Maroon,
    citecolor = Maroon
}

\usepackage{microtype}
\usepackage{verbatim}
\usepackage[amssymb]{SIunits}
\usepackage{tabularx}
\usepackage[dvipsnames]{xcolor}
\usepackage{tikz}
\usepackage{orcidlink}
\usepackage[normalem]{ulem}
\usepackage{xcolor}
\usepackage{soul}

\usepackage{color}

\newcommand{\columbia}{Department of Physics, Columbia University, New York, NY, USA 10027}

\newcommand{\be}{\begin{equation}}
\newcommand{\ee}{\end{equation}}
\newcommand{\bea}{\begin{eqnarray}}
\newcommand{\eea}{\end{eqnarray}}
\newcommand{\beas}{\begin{eqnarray*}}
\newcommand{\eeas}{\end{eqnarray*}}

\newcommand{\rmK}{{\rm K}}

\newcommand{\hatbn}{{\hat{\boldsymbol n}}}

\def\gsim{ \lower .75ex \hbox{$\sim$} \llap{\raise .27ex \hbox{$>$}} }
\def\lsim{ \lower .75ex \hbox{$\sim$} \llap{\raise .27ex \hbox{$<$}} }

\begin{document}
\title{Evidence of galaxy cluster rotation and intrahalo gas motions in the cosmic microwave background}

\author{Samuel~Goldstein}
\email{sjg2215@columbia.edu}
\affiliation{\columbia}

\author{J.~Colin Hill}
\affiliation{\columbia}

\begin{abstract}
\noindent We report the first robust evidence for the rotational kinematic Sunyaev-Zel'dovich (rkSZ) effect: the dipolar signal produced by the Thomson scattering of cosmic microwave background (CMB) photons off intracluster gas in dynamically active systems (including rotations and mergers). By combining CMB intensity and polarization measurements from the \emph{Planck} satellite with spectroscopic member-galaxy redshifts from the Sloan Digital Sky Survey in a sample of 25 X-ray cross-matched, low-redshift ($0.02< z< 0.09)$, massive ($10^{13.9}\lesssim M_{\rm 500c}/M_\odot \lesssim 10^{14.6}$) galaxy clusters, we detect a dipolar rkSZ signature aligned with the estimated angular momentum of each cluster, ruling out a chance fluctuation at 99.98\% confidence (3.6$\sigma$). The significance of this measurement is enhanced by several new methodological improvements for isolating the rkSZ signal from primary CMB fluctuations and noise. The amplitude and shape of the signal are qualitatively consistent with predictions from state-of-the-art hydrodynamical simulations. These results establish a new tool with which to probe the dynamical state of galaxy clusters using CMB data.
\end{abstract}

\maketitle
\paragraph{Introduction---} Galaxy clusters are the largest gravitationally bound structures in the Universe, containing hundreds to thousands of galaxies in a dark matter halo. Understanding the evolution and distribution of gas and dark matter within clusters provides valuable insights into galaxy formation and the growth of structure~\cite{Kravtsov:2012zs}, while the cluster abundance is a potent cosmological observable~\cite[e.g.,][]{Planck:2015lwi}. The scattering of cosmic microwave background (CMB) photons off of free electrons within clusters via the Sunyaev-Zel'dovich (SZ) effect offers a powerful probe of the gas within clusters.

The SZ effect is typically separated into two components: the thermal SZ (tSZ) effect and the kinematic SZ (kSZ) effect~\cite{1969Ap&SS...4..301Z,Sunyaev:1972eq, Sunyaev:1980nv}. The tSZ effect arises from inverse-Compton scattering of CMB photons off energetic electrons in galaxy groups and clusters. This process up-scatters the CMB photons, modifying their frequency distribution and producing a characteristic spectral distortion whose amplitude depends on the electron pressure. In contrast, the kSZ effect is sourced by the Doppler boosting of CMB photons scattering off of electrons with non-zero line-of-sight bulk velocities, making it a unique probe of the momentum of the gas within clusters.

The kSZ effect was first detected in Ref.~\cite{Hand:2012ui} through the large-scale motions of galaxy groups. Since then, the kSZ effect has been detected with increasing significance, enabling precise constraints on astrophysics and the early Universe ~\cite[e.g.,][]{ACTPol:2015teu, Hill:2016dta, AtacamaCosmologyTelescope:2020wtv, Amodeo:2020mmu, Kusiak:2021hai, Hadzhiyska:2024qsl, Krywonos:2024mpb, Lague:2024czc, McCarthy:2024nik, Hadzhiyska:2024ecq, Lai:2025qdw, Hotinli:2025tul, McCarthy:2025brx, Gong:2025ffw}. These measurements primarily probe the kSZ signal sourced by the large-scale motions of halos. However, clusters are dynamic environments with complex small-scale internal gas motions that also contribute to the kSZ signal. High-resolution observations of a few extreme mergers have revealed possible evidence for the kSZ signal sourced by substructure motions~\cite{Sayers:2013ona, Adam:2016abn, Sayers:2018ple}. Furthermore, stacking analyses have found tentative hints of the rotation-induced kSZ effect~\cite{Baxter:2019tze, Matilla:2019yhu}. To date, however, there has been no significant detection of this rotational kSZ (rkSZ) effect, even when considering samples that include merging systems. In this work, ``rotational kSZ effect'' (also known as ``the Baxter effect'') refers broadly to dipolar motion of gas within clusters. In practice, this includes merger-driven flows and coherent rotations, and the distinction between the two is not always clear. Indeed, in the standard hierarchical model of structure formation, dark matter halos acquire angular momentum via a combination of tidal torques, mergers, and accretion of neighboring halos~\cite{Barnes:1987hu, Gardner:2000hj, Maller:2001eu, Vitvitska:2001vw}. The dipolar rkSZ signal is expected to be larger for recent and ongoing mergers compared to coherently rotating clusters~\cite{Cooray:2001vy, Chluba:2002es, Altamura:2023hoe}. Therefore, we adopt this broader definition of rkSZ effect, which is consistent with much of the existing literature (see, \emph{e.g.}, Refs.~\cite{Cooray:2001vy, Chluba:2002es, Altamura:2023hoe}). Nevertheless, future analyses with improved modeling and diagnostics will be necessary for isolating the contribution from coherent cluster rotation specifically.

The rkSZ effect, which was first considered in detail in Refs.~\cite{2001DipT.........1C, Cooray:2001vy, Chluba:2002es}, produces a dipolar pattern that is aligned with the rotation direction, i.e., the angular momentum, of the intracluster medium (ICM), as electrons on one side of the halo move toward the observer and those on the other recede. Simulations predict rkSZ dipole amplitudes of order $\sim 10-100~\mu$K in galaxy clusters~\cite{Baldi:2018eje, Altamura:2023hoe}, depending on a range of properties, including the cluster's mass and dynamical state. For example, Refs.~\cite{Altamura:2023hoe, Monllor-Berbegal:2024osg} found that the rkSZ signal in clusters with recent or ongoing mergers can be over a factor of two larger than in relaxed clusters. Measuring this rkSZ signal would shed light on the properties of the ICM, galaxy formation, and the growth of structure~\cite{Dupke:2002nh, Lau:2013ora, Montero-Dorta:2020dnd}.

In this \emph{Letter}, we present the first robust observational evidence for the rkSZ effect. Building upon the oriented-stacking analysis of~\citet{Baxter:2019tze}, which found a modest ($\sim 2\sigma$) hint for a dipolar kSZ signal in a sample of 13 clusters, we perform an oriented stack on a larger sample of 25 clusters and detect a significant ($3.6\sigma$) dipole. Our improved signal-to-noise arises not only from using a larger cluster sample, but also from several methodological advances, which can be applied to future kSZ analyses. These results open a new window for using CMB observations to probe the dynamical state of galaxy clusters.  The Supplemental Material (SM) contains extensive validation of our measurement.

\noindent\emph{Conventions:} We assume a flat $\Lambda$CDM cosmology based on the \emph{Planck}+ACT (P-ACT) analysis~\cite{AtacamaCosmologyTelescope:2025blo}:\footnote{\href{https://lambda.gsfc.nasa.gov/data/act/chains/lcdm/p-actbase_lcdm_class.tar.gz}{https://lambda.gsfc.nasa.gov/data/act/chains/lcdm/p-actbase\_lcdm\_class.tar.gz}} $\Omega_{\rm cdm}h^2=0.1195$, $\Omega_{\rm b}h^2=0.0225$, $h=0.6781$, $\ln(10^{10}\,A_s)=3.0447$, $n_s=0.9723$, and $\tau=0.057$. Here, $A_s$ and $n_s$ are the amplitude and spectral index of primordial scalar perturbations, $\tau$ is the optical depth to reionization, $h \equiv H_0/100~{\rm km/s/Mpc}$ is the dimensionless Hubble constant, and $\Omega_{\rm cdm}h^2$ and $\Omega_{\rm b}h^2$ are the present-day physical cold dark matter and baryon densities, respectively. We define $R_{\rm 500c}$ as the radius within which the mean enclosed density of a halo is 500 times the critical density at the halo redshift, and $M_{\rm 500c}$ as the mass enclosed within this radius. We use $r$ ($R$) to denote the 3D (projected 2D) radial cluster-centric distance.

\paragraph{Theory---} The kSZ-induced relative CMB temperature shift due to a galaxy cluster in a direction $\hatbn$ on the sky can be expressed as a line-of-sight integral, 
\begin{equation}\label{eq:kSZ_dipole}
    \frac{\Delta T_{\rm kSZ}(\hatbn)}{T_{\rm CMB} }=-\frac{\sigma_T}{c}\int\limits_{\rm LoS}dl\, n_e\,{\bm v}_{\rm pec}\cdot \hatbn,
\end{equation}
where $T_{\rm CMB}=2.7255~\rmK$ is the CMB monopole temperature~\cite{2009ApJ...707..916F}, $n_e$ is the free-electron number density, ${\bm v}_{\rm pec}\cdot \hatbn$ is the peculiar velocity of the electrons along the line-of-sight, $\sigma_T$ is the Thomson scattering cross section, and $c$ is the speed of light. 

The kSZ signal is difficult to separate from primary CMB fluctuations because it preserves the CMB's blackbody spectral energy distribution (SED), but external velocity information enables robust kSZ inference via cross-correlation, as other signals in the microwave sky do not trace the velocity field~\cite[e.g.,][]{Hand:2012ui,AtacamaCosmologyTelescope:2020wtv}. A cluster moving coherently relative to the CMB produces a \emph{monopole}-like kSZ signal at the cluster location. In contrast, a cluster rotating about an axis that is not parallel to the line-of-sight generates a \emph{dipole}-like kSZ signal, as one side of the cluster moves toward the observer and the other recedes. We also note that similar dipoles arise in recent and ongoing mergers -- indeed, such mergers are the origin of the rotational gas motions within halos. All of these contributions play a role in the rkSZ signal explored in this work.

Theoretical and simulation-based studies suggest that the rkSZ dipole in massive galaxy clusters is of order $\sim 10-100~\mu$K~\cite{Cooray:2001vy, Chluba:2002es, Baldi:2018eje, Altamura:2023hoe}. Detecting this small signal in nearby clusters is challenging because the microwave sky is dominated by the primary CMB on these scales, making it necessary to average over multiple clusters. Furthermore, na\"ively stacking CMB maps at cluster locations would wash out the dipole unless the maps are aligned with an external estimate of the cluster's rotation direction, which is difficult to obtain~\cite{Manolopoulou:2016ozk, Tang:2025lmm, Castellani:2025kon}. Following Ref.~\cite{Baxter:2019tze}, we estimate a cluster's rotation direction using spectroscopic redshifts of its member galaxies. Although the misalignment between the true gas rotation direction and the galaxy-inferred rotation direction can suppress the rkSZ signal by more than a factor of two~\cite{Altamura:2023hoe}, a statistically significant dipole aligned with the galaxy-inferred direction constitutes strong, foreground-robust evidence for the rkSZ effect.

\paragraph{Datasets and methodology---} To isolate the small rkSZ signal, we first combine multi-frequency CMB observations to build a map with enhanced sensitivity to the rkSZ. We use the internal linear combination (ILC) method~\cite{COBE_1992, WMAP:2003cmr, Tegmark:2003ve, Eriksen:2004jg} to form the minimum-variance linear combination of a set of single-frequency maps with unit response to the CMB blackbody SED. Here, we use needlet ILC (NILC)~\cite{Delabrouille:2008qd,Remazeilles:2010hq,McCarthy:2023hpa}, a variant of ILC that computes the weights on a frame of needlets~\cite{doi:10.1137/040614359}, which have compact support in real and harmonic space. NILC is well suited for CMB analyses because Galactic foregrounds are highly anisotropic in real space, while extragalactic foregrounds are more naturally separated in harmonic space due to their statistical isotropy.  Although standard NILC already suppresses much of the tSZ signal, we use \emph{constrained} NILC~\cite{Chen:2008gw, Remazeilles:2010hq} to ``deproject,'' i.e., exactly remove contributions from the non-relativistic tSZ effect.\footnote{We neglect the relativistic SZ effect~\cite{Itoh:1997ks, Challinor:1997fy, Sazonov:1998ae}, which is not aligned with cluster rotation and is far smaller than the minor residual tSZ signal in non–tSZ-deprojected maps (see the SM).} The tSZ effect does not produce a dipole correlated with cluster rotation, but it would contribute excess noise at the cluster locations. In the SM, we show that a significant dipole is present in the oriented cluster stack even without tSZ deprojection.

We construct NILC $T$ maps using \texttt{pyilc}~\cite{McCarthy:2023hpa}\footnote{\href{https://github.com/jcolinhill/pyilc/}{https://github.com/jcolinhill/pyilc/}}\footnote{We construct our own NILC maps, rather than using the \textsc{SMICA} map employed in Ref.~\cite{Baxter:2019tze}, to assess the robustness of our results to variations in the NILC pipeline. In the SM, we repeat our analysis with the \texttt{SMICA-noSZ} map and obtain similar results.}, closely following the methodology in Refs.~\cite{McCarthy:2023hpa, McCarthy:2023cwg, McCarthy:2024ozh, Goldstein:2024mfp}. We use the \emph{Planck} NPIPE (PR4)~\cite{Planck:2020olo} temperature maps\footnote{We do not include the latest high-resolution CMB measurements from ACT~\cite{ACT:2023wcq, AtacamaCosmologyTelescope:2025vnj} because most of our clusters lie outside of the ACT footprint. Furthermore, the clusters are very nearby ($z \lesssim 0.1$) and therefore span relatively large angular scales where the ACT data are limited by atmospheric noise.} at 30, 44, 70, 100, 143, 217, 353, and 545 GHz, which we reconvolve to a common Gaussian beam with a full-width half-maximum (FWHM) of 5 arcminutes. Before performing the component separation, we fill in regions with bright point sources and Galactic emission using a diffusive inpainting algorithm to prevent these regions from contributing to the NILC weights. The final component-separated map is a full-sky, tSZ-deprojected blackbody temperature map. 

For our rkSZ effect measurement, the dominant source of noise in the NILC map is the primary CMB. To suppress this noise, while preserving the rkSZ signal, we exploit the (partial) correlation between the primary CMB temperature and E-mode polarization. Specifically, since the E-modes are effectively uncorrelated with the rkSZ effect,\footnote{We neglect polarized contributions to the SZ effect, which are significantly below the level of sensitivity of the datasets used here~\cite[e.g.,][]{Sunyaev:1980nv, Sazonov:1999zp, Chluba:2002es}.} we construct an ``E-mode subtracted" temperature map, $\tilde{T}$, defined in harmonic space by
\begin{equation}\label{eq:Emode_subtraction}
    a_{\ell m}^{\tilde{T}}=a_{\ell m}^{T}-\frac{{C}_{\ell}^{TE;~{\rm obs}}}{C_{\ell}^{EE;~{\rm obs}}}\;a_{\ell m}^{E}.
\end{equation}
Here, $a_{\ell m}^T$ and $a_{\ell m}^E$ are the spherical harmonic coefficients of the blackbody temperature and polarization maps, and $C_{\ell}^{EE;~{\rm obs}}$ and $C_{\ell}^{TE;~{\rm obs}}$ are the \emph{observed} E-mode auto-power spectrum and temperature-E-mode cross-power spectrum, respectively. 

For the E-mode data, we use the \emph{Planck} PR3 \textsc{SMICA} map~\cite{Planck:2018yye}. Before computing the spherical harmonic coefficients in Eq.~\eqref{eq:Emode_subtraction}, we apply a common analysis mask defined by the intersection of the inpainting mask and the \emph{Planck} PR3 polarization common confidence mask~\cite{Planck:2018yye}, apodized using the \textsc{Namaster} C2 cosine apodization with a scale of 30 arcminutes~\cite{Grain:2009wq, Alonso:2018jzx}.  We set $C_{\ell}^{TE;~{\rm obs}} = C_{\ell}^{TE}$ computed at the fiducial cosmology because the measured $TE$ cross-spectrum does not contain an additive noise bias. In contrast, the $EE$ auto-spectrum contains a large additive noise bias, thus we estimate $\hat{C}_{\ell}^{EE}$ directly from the data and smooth it with a Savitzky-Golay filter~\cite{1964AnaCh..36.1627S} with window length five and order three.

We subtract E-modes across all multipoles between $50\leq \ell \leq 2000$, reducing the large-scale primary-CMB contribution to the NILC $T$ map auto-spectrum by up to 20\%, as discussed in the SM. To our knowledge, this is the first application of this procedure, which was originally proposed in Ref.~\cite{Frommert:2008qh}, to data. As a final step, we apply a harmonic-space filter to our $\tilde{T}$ map that removes modes with $\ell \lesssim 100$ and $\ell \gtrsim 3000$, thus removing modes dominated by the primary CMB and instrumental noise, respectively. In the SM, we show that the combination of E-mode subtraction and harmonic-space filtering increases the stacked rkSZ dipole significance by $\sim 0.5$–$1.0\sigma$; nevertheless, a clear dipole signal remains even without these optimizations.

We construct a galaxy cluster catalog with estimated rotation directions following~\citet{Tang:2025lmm}, which defines the sky-projected rotation direction as the direction that maximizes the redshift difference between member galaxies split into two hemispheres. We use the Friends-of-Friends (FoF) group catalog from~\cite{Tempel:2017dhe}, based on Sloan Digital Sky Survey (SDSS) Data Release 12 (DR12) observations. This catalog consists of 88,662 groups/clusters with a total of 287,245 member galaxies. Crucially, this catalog splits clusters into potential substructures, allowing rotation directions to be estimated per substructure and thereby boosting sensitivity to rotational, rather than merger-driven, kSZ signals. To mitigate optical cluster selection biases, we cross-match with the MCXC-II X-ray cluster meta-catalog~\cite{Sadibekova:2024jnb}, requiring the optical center to lie within a projected distance of 10 arcminutes from the X-ray center, based on the angular resolution of our NILC maps\footnote{The optical center refers to the group center of the FoF group catalog~\cite{Tempel:2017dhe}, defined as the unweighted geometric center of all galaxies within the group~\cite{Tempel:2016gdu}. The X-ray center is the location of peak X-ray luminosity from the MCXC-II catalog. As shown in the SM, we find similar results using the optical center.} and a redshift difference of $|\Delta z|\leq 0.003$. Following previous analyses~\cite{Manolopoulou:2016ozk, Castellani:2025kon}, we restrict to groups/clusters with at least 50 member galaxies within a projected distance of $3\,R_{\rm 500c}$ to the X-ray center. Finally, we keep only clusters with $\geq90\%$ of their area unmasked in a square cutout of side length $2\, R_{\rm 500c}$,\footnote{We exclude pixels with a mask value below 0.5 from the stack.} yielding a sample of 54 clusters. The typical cutout map spans roughly 20-40 arcmin on a side.

To estimate the rotation direction of each cluster, we project all member galaxies onto the tangent plane centered on the X-ray center. We then divide the projected distribution into left and right hemispheres using a ray through the origin at an angle $\theta_{\rm trial}$, defined clockwise from the vertical axis (see Fig.~\ref{fig:coord_system_def} in the SM). For each $\theta_{\rm trial}$, we compute the difference in the mean galaxy redshift between the left and right hemispheres. Scanning over $0^\circ \leq \theta_{\rm trial} < 360^\circ$ in steps of $5^\circ$, we identify the rotation direction $\theta_{\rm rot}$ as the angle that \emph{maximizes} the redshift difference between galaxies in the two hemispheres, corresponding to preferential blueshifts (redshifts) on the right (left) of the rotation direction vector. The maximum absolute redshift difference, $|\Delta z_{\rm dip}^{\rm max}|$, is a proxy for the rotational velocity. To remove clusters with weak or poorly determined rotation directions, we compute the correlation coefficient between the measured redshift-difference curve and a sinusoidal template with amplitude $|\Delta z_{\rm dip}^{\rm max}|$ and maximum at $\theta_{\rm rot}$. We keep clusters with correlation coefficients $\geq 0.8$, yielding a final sample of 25 clusters.

Using this final cluster sample, we extract cutouts from the filtered NILC $\tilde{T}$ map centered on each cluster's X-ray position. We rotate each cutout from Galactic to Equatorial coordinates and orient it such that the positive vertical axis aligns with the cluster’s estimated rotation direction, with the right-hand side rotating toward the observer. To reduce scatter in the stack, we rescale each cutout to a square of side length $2R_{\rm 500c}$ centered on the cluster. Finally, we compute a weighted-average stack, assigning each cluster the following approximate signal-to-noise weight:
\begin{equation}\label{eq:fid_weights}
    w=\frac{M_{\rm 500c}\, |\Delta z_{\rm dip}^{\rm max}| N_{\rm mem}}{\sigma^2_{\tilde{T}}(M_{\rm 500c}, z)}.
\end{equation}
Here, the signal scales as $n_e v_{\rm rot} \propto M_{\rm 500c} |\Delta z_{\rm dip}^{\rm max}|$, and we assume that the variance on the rotation direction estimate is inversely proportional to the number of member galaxies, $N_{\rm mem}$. Finally, $\sigma_{\tilde{T}}^2(M_{\rm 500c}, z)$ is a Monte Carlo estimate of  the variance in the measured dipole amplitude for a cutout of size $R_{\rm 500c}$ at redshift $z$, as detailed below.

Following Ref.~\cite{Baxter:2019tze}, we estimate the dipole amplitude, $\langle D \rangle$, as \emph{half} the difference in the mean temperature between the right and left hemispheres over a particular set of pixels.\footnote{This differs from the definition in Ref.~\cite{Baxter:2019tze} by a factor of 1/2.} Motivated by the simulated rkSZ profiles in Ref.~\cite{Altamura:2023hoe}, we use pixels with $|x/R_{\rm 500c}|\geq 0.1$ and $R/R_{\rm 500c}<0.75$. As shown in the SM, the significance of the measured dipole is largely insensitive to the size and shape of this pixel-space filter.

\begin{figure}[!t]
\centering
\includegraphics[width=0.99\linewidth]{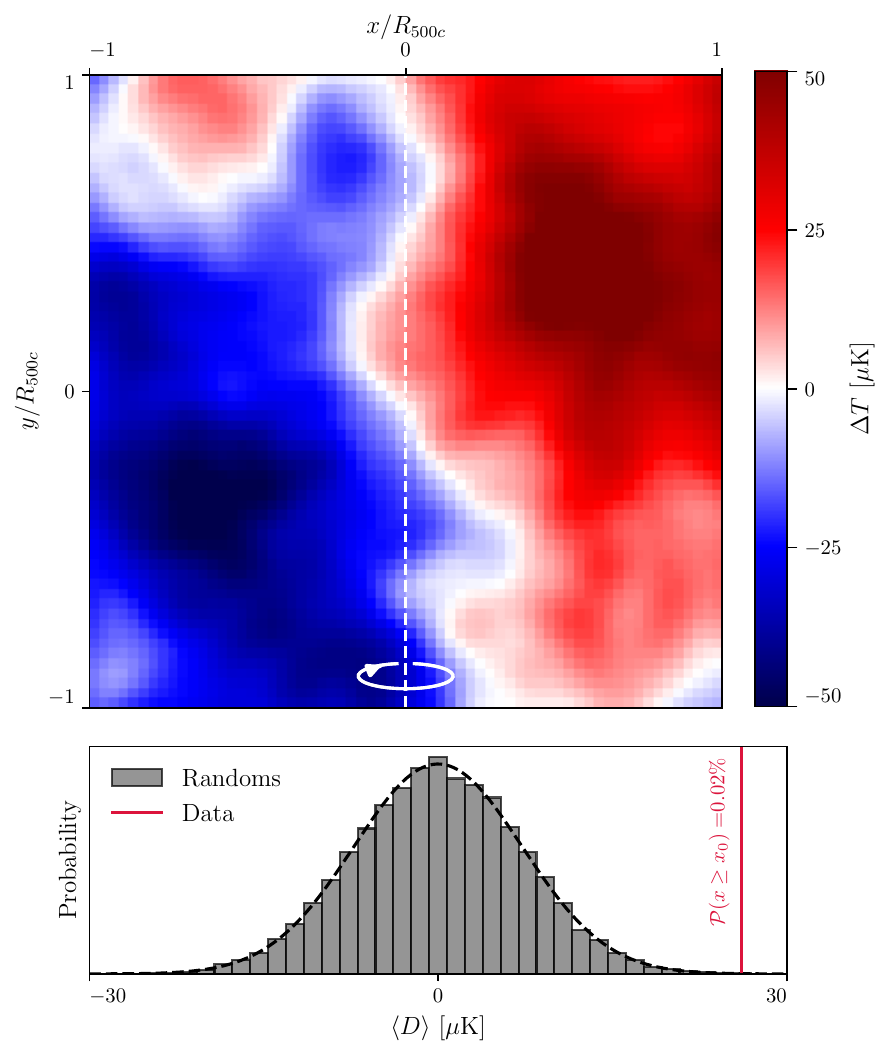}
\caption{\emph{Top:} oriented and weighted stack of filtered and cleaned \emph{Planck} CMB temperature data on 25 galaxy clusters. The cutouts are rescaled in units of $R_{\rm 500c}$ and aligned with the estimated cluster rotation direction, with the right side ($x \geq 0$) rotating towards the observer, as indicated by the white arrow. A dipole signal is clearly visible along the expected rotation direction. \emph{Bottom:} significance of the dipole estimated from the oriented stack (red line) compared with the noise distribution obtained by repeating the weighted stack 30,000 times at random locations. The measured dipole amplitude exceeds 99.98\% of the random realizations, indicating strong evidence ($3.6\sigma$) of cluster rotation. The black dashed line shows the best-fit Gaussian to the noise.}
\label{fig:stacked_rkSZ}
\end{figure}

To assess the dipole significance, we generate 30,000 random stacks. Each stack uses 25 random positions uniformly drawn from the intersection of the SDSS group catalog footprint and our analysis mask. For each position, we extract a square cutout of side length $2R_{\rm 500c}$ (set by the cluster's mass and redshift), require at least 90\% of the pixels are unmasked, and measure the dipole using the same filter as in the data. For each cluster, we compute ${\sigma^2_{\tilde{T}}(M_{\rm 500c}, z)}$ as the variance of the dipole amplitudes across the 30,000 randoms, which captures the likelihood of obtaining a dipole from chance fluctuations for that cutout size. Using these variances, we compute the cluster weights (Eq.~\eqref{eq:fid_weights}) and perform the weighted stack on the data and all 30,000 random realizations. We then directly quantify the dipole significance by comparing the measured dipole amplitude to the distribution of randoms. This procedure, based on Ref.~\cite{Baxter:2019tze}, neglects sources of cluster-correlated noise. The main sources of cluster-correlated noise that produce a dipole are CMB cluster lensing and the moving lens effect, but these are subdominant compared to the primary CMB and instrumental noise over the range of scales and redshifts used here~\cite{Hotinli:2018yyc, Beheshti:2024dxw,Hotinli:2024tjb}. Furthermore, non-dipolar signals such as the bulk kSZ, CIB, and the relativistic SZ, which will not bias the signal, could contribute excess noise at cluster locations. In the SM, we confirm that cluster-correlated noise is subdominant by repeating our stack on the same set of clusters with randomized rotation angles. Adopting this noise model, we find evidence for a dipole at $\sim 2.9\sigma$, although we note that there are several caveats to this analysis, which we discuss in detail in the SM. Future rkSZ analyses of higher-redshift clusters ($z\gtrsim0.5$) will need to carefully model, and potentially subtract, these contributions.

Finally, we compute the azimuthally-averaged temperature profile within each hemisphere in six radial bins from $-1 \leq R/R_{\rm 500c} \leq 1$, with the covariance estimated from the 30,000 random stacks. We compare the measured profile to a simplified theoretical template in which each cluster undergoes solid-body rotation with $v_{\rm rot}=1500~{\rm km/s}$ at $0.2\,r_{\rm 500c}$ and electrons follow the “AGN Feedback’’ density profile of Ref.~\cite{Battaglia:2016xbi}.\footnote{We choose the rotational velocity such that the amplitude of our rkSZ effect profile is $\sim 25~{\mu {\rm K}}$ to match the hydrodynamical simulation predictions from Ref.~\cite{Altamura:2023hoe}.} We compute the weighted average of the beam-convolved theoretical profiles across our cluster sample. We stress that this simplified model is used only to study qualitative features of the rkSZ signal, as the detection significance in this work is insufficient to extract precise constraints on the gas’s rotational motion. Indeed, for the dynamically disturbed systems in our sample, the spherically averaged density profile does not account for substructure or merging components, which can enhance the electron density in the cluster outskirts. Therefore, the electron density and velocity profiles adopted here are purely used for qualitative purposes, similar to Ref.~\cite{Baxter:2019tze}. To derive physically meaningful results from the overall shape and amplitude of the profile, one needs a dedicated model of the gas distribution and merger properties of the cluster sample, which is significantly beyond the scope of this work. Additional details on the profile measurement and model are provided in the SM.

\paragraph{Results---}  

The top panel of Fig.~\ref{fig:stacked_rkSZ} shows the oriented, weighted stack of the filtered \emph{Planck} data on our cluster sample. A dipole with amplitude $\approx 25 \, \mu\rm K$ is visually apparent and aligned with the estimated rotation direction (vertical dashed line). The bottom panel compares the measured dipole amplitude with the distribution of randoms. The black dashed line shows a Gaussian fit to the resulting noise distribution. Using this Gaussian approximation,\footnote{A direct significance estimate from the randoms also gives $3.6\sigma$.} the dipole exceeds 99.98\% of random realizations, corresponding to $3.6\,\sigma$ evidence for the rkSZ effect. The dipole is not perfectly aligned with the estimated rotation axis, but we caution against over-interpreting this result for our relatively small cluster sample, as such misalignment can arise from differences between the true and estimated rotation directions, cluster substructure, and residual primary CMB or instrumental noise.

\begin{figure}[!t]
\centering
\includegraphics[width=0.99\linewidth]{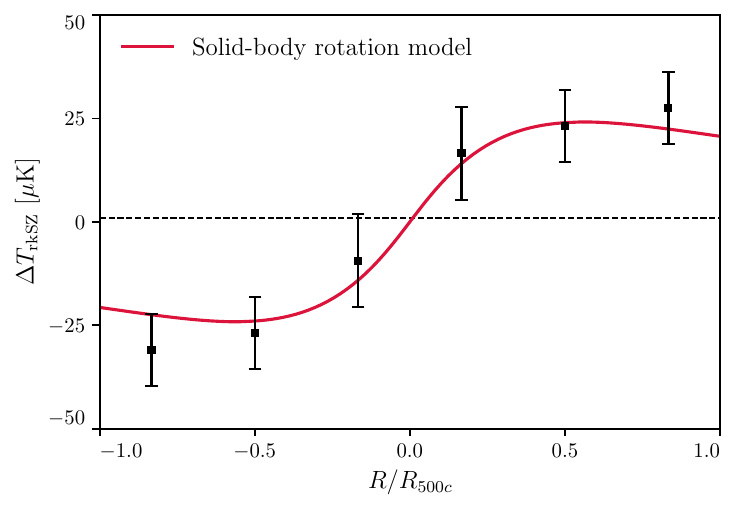}
\caption{Azimuthally-averaged rkSZ effect profile derived from our weighted stack of 25 clusters. The red curve shows a theoretical prediction based on a solid-body rotation model and the Battaglia~\cite{Battaglia:2016xbi} electron number density profile. We find significant evidence for a non-zero amplitude, $A_{\rm rkSZ}=1.05\pm 0.32$. Note that the data points are highly correlated.}
\label{fig:rkSZ_profile}
\end{figure}

Fig.~\ref{fig:rkSZ_profile} shows the azimuthally-averaged rkSZ profile in each hemisphere as a function of radial distance from the cluster center. The black points are measured from the oriented stack in Fig.~\ref{fig:stacked_rkSZ}, with significant correlations across bins. The red curve shows the prediction of our simple theoretical model, rescaled by an amplitude, $A_{\rm rkSZ}$. Fitting for this amplitude yields significant evidence for a non-zero signal, $A_{\rm rkSZ}=1.05\pm0.32$ (68\% confidence limit).\footnote{The fact that $A_{\rm rkSZ}\approx 1$ should not be interpreted as validation of our theoretical model. $A_{\rm rkSZ}$ is degenerate with the rotational velocity, which we fix to have a dipole amplitude of order 25$~\mu{\rm K}$. To extract physical meaning from $A_{\rm rkSZ}$, one would need direct estimates of each cluster's rotational velocity and would also need to account for the misalignment between the inferred and true rotation directions, as well as projection effects.}

Although a detailed analysis of the rkSZ profile would require a higher-significance measurement, it is instructive to compare our results to the hydrodynamical simulations in Ref.~\cite{Altamura:2023hoe}. Their predicted dipole amplitude ($\sim 25~\mu$K) and profile shape, when stacking on the galaxy-inferred rotation directions, are broadly consistent with our measurements. Our profile is perhaps somewhat more radially extended than their simulation results, although this can be partially attributed to beam convolution. Overall, our measurement is qualitatively consistent with Ref.~\cite{Altamura:2023hoe}.

Having found evidence for a statistically significant dipole in our component-separated maps that is correlated with the inferred cluster rotation directions, we discuss alternative explanations for the signal; further robustness and null tests can be found in the SM. First, the Doppler boosting of dust emission from rotating galaxies within a cluster leads to a dipole that is aligned with the cluster rotation direction; however, galaxies in massive clusters at low redshift have very little dust emission, and furthermore any such contribution is suppressed in our component-separated maps~\cite{Maniyar:2022lkv}. In the SM, we explicitly demonstrate that no dipole is seen if we repeat our oriented stack on the dust-dominated \emph{Planck} 545 GHz map. Second, gravitational lensing of the CMB by galaxy clusters produces a dipole that preserves the CMB blackbody SED~\cite{Seljak:1999zn, Holder:2004rp, Dodelson:2004as, Baxter:2014frs}. However, this dipole is uncorrelated with the cluster's rotation direction (in fact, it is correlated with the local primary CMB gradient), and thus it will average out in our measurement. Moreover, the geometric CMB lensing kernel is very small for nearby clusters. Finally, the moving lens effect can imprint a dipolar pattern in the CMB, but its expected amplitude is an order of magnitude (or more) lower than our signal and its direction is aligned with the transverse motion of the cluster~\cite{1983Natur.302..315B, Hotinli:2018yyc, Baxter:2019tze}, and thus it will average out in our stack. Overall, we conclude that the statistically significant dipole in our oriented stack is due to the kSZ effect sourced by the internal motion of gas within galaxy clusters, including ongoing and recent mergers.

\paragraph{Conclusions---} In this \emph{Letter}, we present robust evidence ($3.6\sigma$) for a dipole in CMB temperature maps centered on nearby clusters that is correlated with their rotation directions, estimated from the motion of galaxies within each cluster. We attribute this signal to the rotational kSZ effect, i.e., the Doppler boosting of CMB photons scattering off of dipolar motion of electrons within the ICM. Our measured dipole, whose signal-to-noise is enhanced by several methodological improvements designed to minimize contributions from the primary CMB and noise, is qualitatively consistent with predictions from hydrodynamical simulations. 

In the near future, applying these techniques to larger cluster samples and higher-resolution CMB maps will enable precise characterization of the dynamical state of the gas within galaxy clusters. Full-sky, low-noise CMB polarization measurements from experiments such as\emph{LiteBIRD} would also improve rkSZ searches based on the methodology used in this work~\cite{LiteBIRD:2022cnt}. Existing X-ray and tSZ cluster catalogs contain $\sim 100-1000\times$ the number of clusters analyzed here~\cite{Bulbul:2024mfj,ACT:2025fip,SPT:2014wbo,Planck:2015lwi}, thus, with sufficient optical spectroscopy and high-resolution CMB data, one can expect a sizable increase in signal-to-noise. These improvements will enable rkSZ measurements in lower-mass halos~\cite{Matilla:2019yhu}, and, crucially, allow comparisons across clusters split by dynamical state, helping distinguish coherent rotation from merger-driven flows and other sources of dipolar structure, such as gas sloshing.  Amongst other applications, this will enable precise constraints on rotational support in the ICM, as may play a role in the hydrostatic bias in cluster mass estimates~\cite[e.g.,][]{Nelson:2013hwa,Shi:2015fua,Gianfagna:2022pbv}, the non-observation of cooling flows in some relaxed clusters~\cite[e.g., Abell 2029,][]{Lewis:2002kp}, and other important aspects of ICM physics.

\vspace{5pt}

\paragraph{Data availability statement---}
Our data products and analysis pipeline are publicly available as part of the \texttt{aardvarkSZ} package at \href{https://github.com/samgolds/aardvarkSZ/}{https://github.com/samgolds/aardvarkSZ/}.

\paragraph{Acknowledgments---}

\noindent We thank Jens Chluba, Eric Gawiser, Mat Madhavacheril, Michael McDonald,  Blake Sherwin, Beatriz Tucci, and Peng Wang for useful discussions. SG and JCH acknowledge support from NSF grant AST-2307727.  JCH also acknowledges support from NASA grant 80NSSC23K0463 [ADAP] and the Sloan Foundation.  The authors acknowledge the Texas Advanced Computing Center (TACC) at The University of Texas at Austin for providing computational resources that have contributed to the research results reported within this paper. We acknowledge computing resources from Columbia University's Shared Research Computing Facility project, which is supported by NIH Research Facility Improvement Grant 1G20RR030893-01, and associated funds from the New York State Empire State Development, Division of Science Technology and Innovation (NYSTAR) Contract C090171, both awarded April 15, 2010. This research has made use of the M2C Galaxy Cluster Database, constructed as part of the ERC project M2C (The Most Massive Clusters across cosmic time, ERC-Adv grant No. 340519).

\bibliographystyle{apsrev4-1}
\bibliography{biblio}

\onecolumngrid
\clearpage
\appendix

\renewcommand{\thesection}{\Alph{section}} 
\setcounter{section}{0} 
\setcounter{secnumdepth}{2}

\begin{figure}[!t]
\centering

\begin{minipage}[c]{0.49\linewidth}
    \centering
    \includegraphics[width=\linewidth]{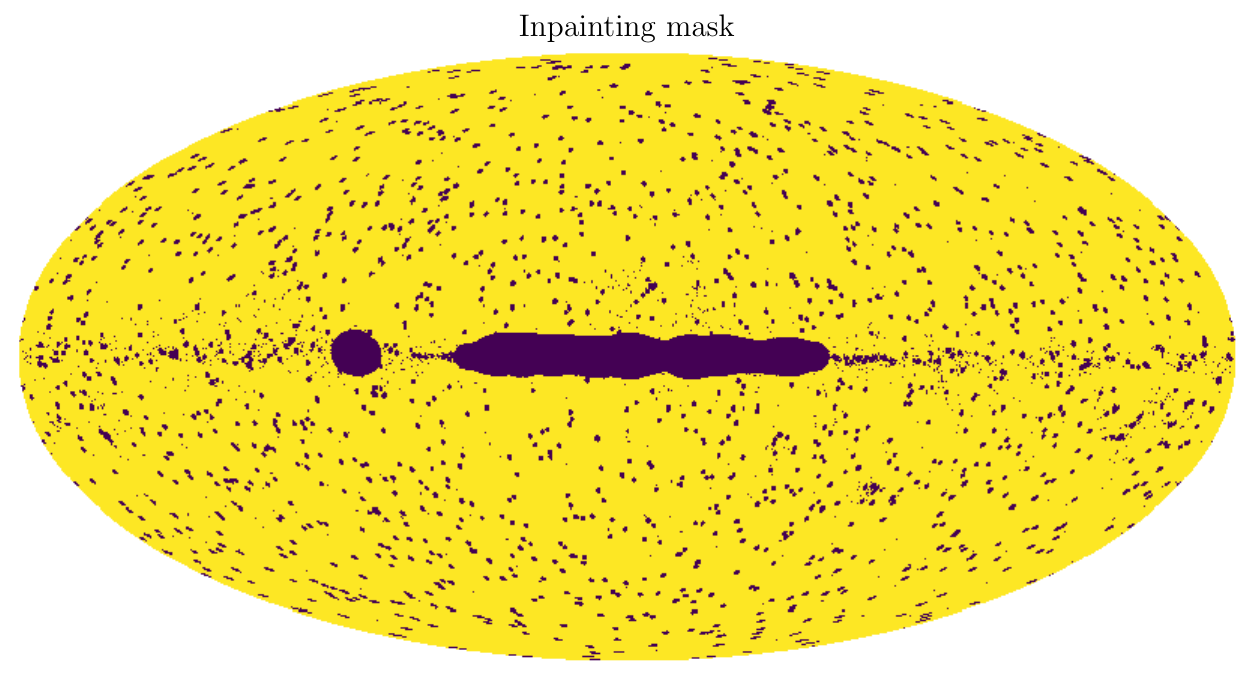}
\end{minipage}
\hfill
\begin{minipage}[c]{0.49\linewidth}
    \centering
    \includegraphics[width=\linewidth]{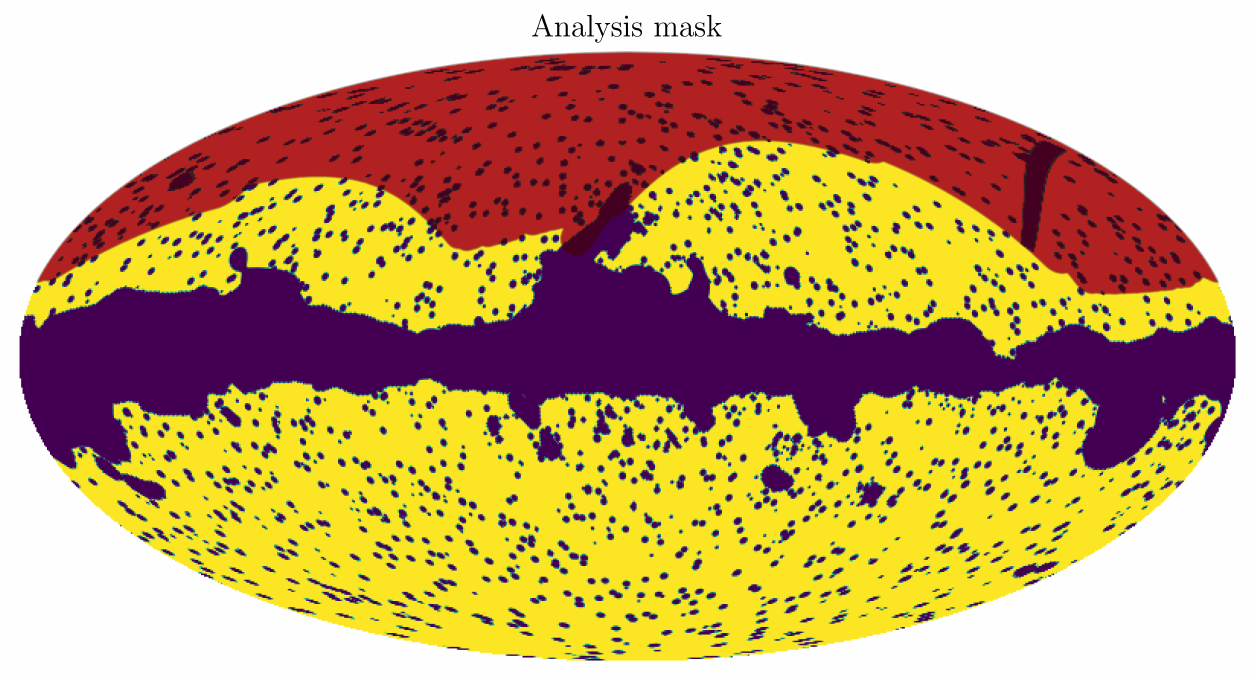}
\end{minipage}

\vspace{0.3cm}
\begin{minipage}[c]{0.49\linewidth}
    \centering
    \includegraphics[width=\linewidth]{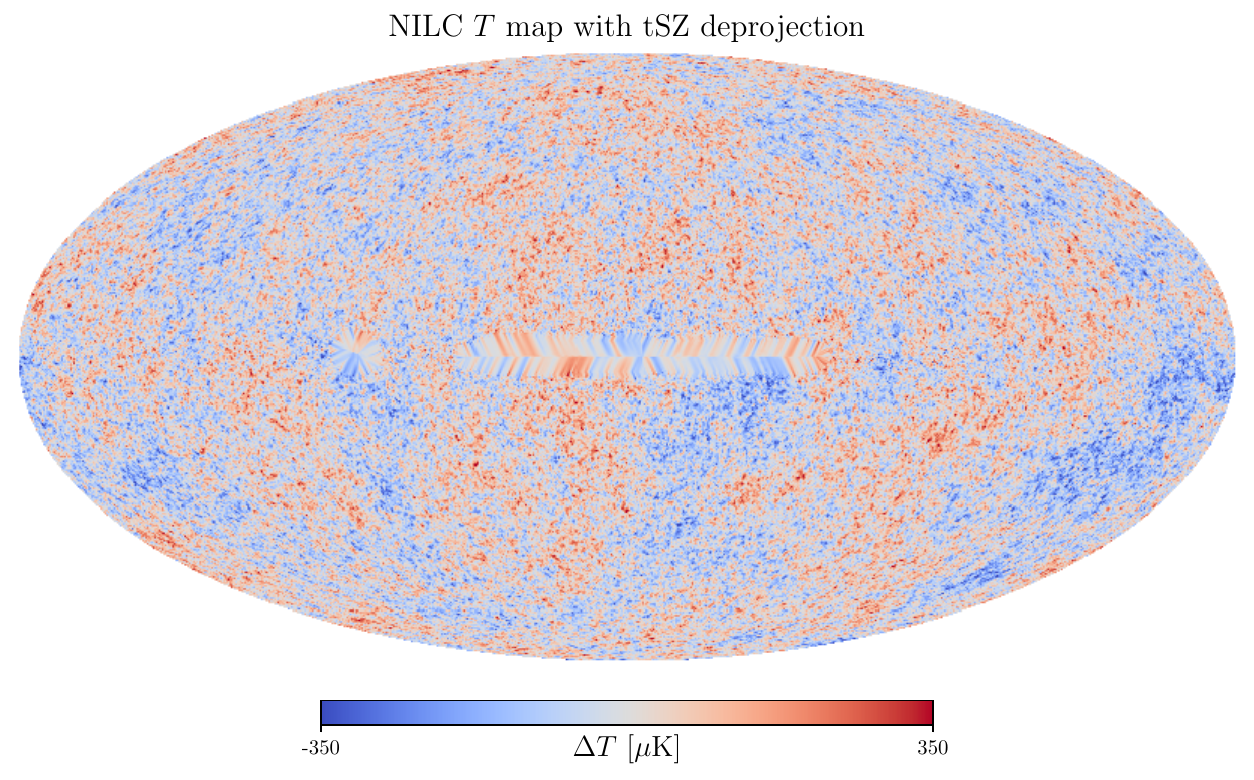}
\end{minipage}
\hfill
\begin{minipage}[c]{0.49\linewidth}
    \centering
    \includegraphics[width=\linewidth]{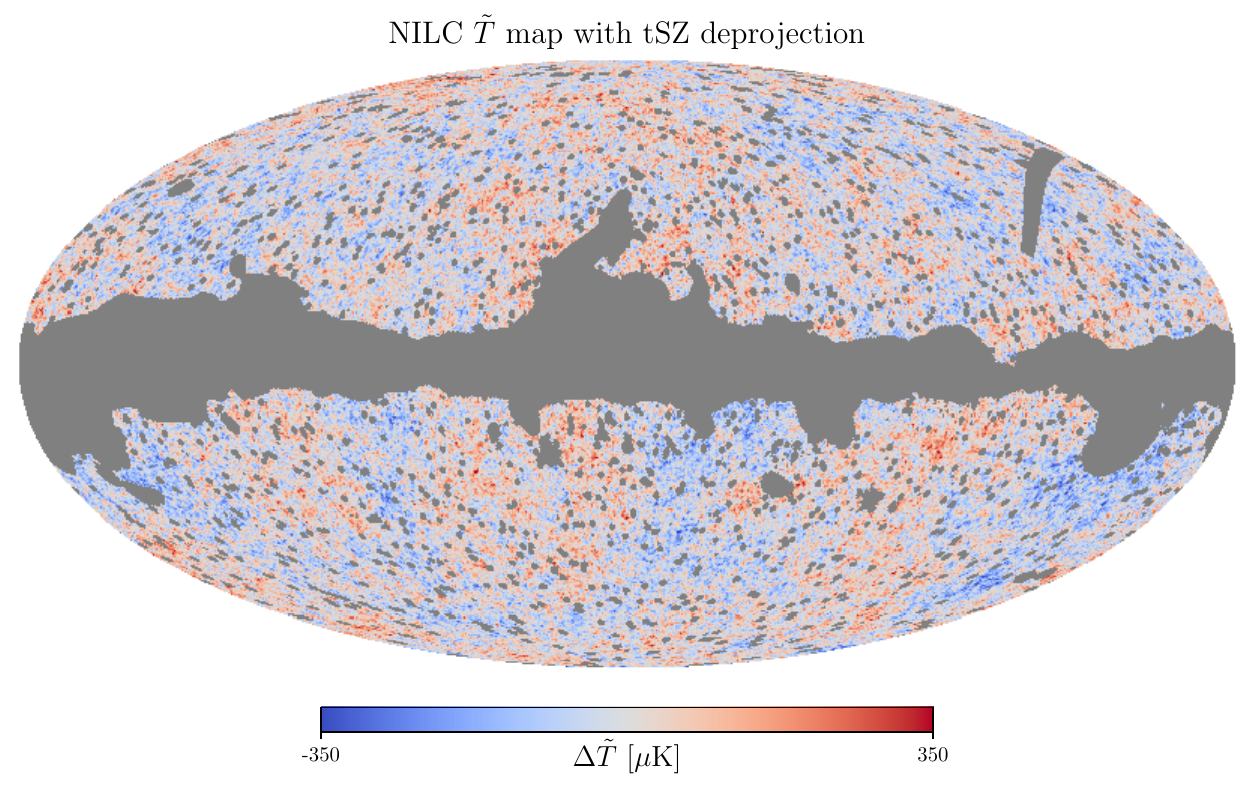}
\end{minipage}

\caption{Masks and main CMB maps used in this work. The top-left panel shows the preprocessing mask used for inpainting the multi-frequency maps prior to component separation. It is constructed from the union of a Galactic plane mask and the \emph{Planck} LFI and HFI point-source masks. The top right panel shows the analysis mask, defined as the union of the inpainting mask and the \emph{Planck} polarization confidence mask, and subsequently apodized with a 30-arcminute {\tt NaMaster} C2 apodization. The red shaded region indicates the subset of the SDSS footprint from which our cluster catalog is drawn. We use the union of this footprint and the analysis mask when performing random stacks when determining the significance of our measurements. The bottom-left shows the full-sky tSZ-deprojected blackbody temperature NILC map constructed from multi-frequency \emph{Planck} intensity measurements with {\tt pyilc}. The bottom-right panel shows the E-mode-cleaned tSZ-deprojected blackbody temperature NILC map, $\tilde{T}$, which we use for our fiducial analysis. As discussed in the text, the $\tilde{T}$ map is constructed from maps that have already been masked to ensure numerical stability in the spherical-harmonic transforms.}
\label{fig:masks_maps_grid}
\end{figure}

\section{CMB maps and processing}

\noindent In this section, we provide additional details regarding the map-making and preprocessing choices used to extract the rotational kinematic Sunyaev–Zel’dovich (rkSZ) signal from multi-frequency cosmic microwave background (CMB) observations. 

\begin{figure*}[t]
\centering

\begin{minipage}[!t]{0.35\textwidth}
\centering
\begin{tabular}{|c|c|c|c|}
\hline
$I$ & $\ell_{\rm peak}$ & FWHM [degrees] & Excluded freqs. \\\hline
0  & 0    & 91.8 & -- \\\hline
1  & 100  & 35.6 & -- \\\hline
2  & 200  & 25.2 & -- \\\hline
3  & 300  & 20.6 & -- \\\hline
4  & 400  & 14.1 & -- \\\hline
5  & 600  & 10.3 & -- \\\hline
6  & 800  & 8.26 & 30~GHz \\\hline
7  & 1000 & 6.31 & 30, 44~GHz \\\hline
8  & 1250 & 6.11 & 30, 44~GHz \\\hline
9  & 1400 & 4.74 & 30, 44~GHz \\\hline
10 & 1800 & 3.55 & 30, 44~GHz \\\hline
11 & 2200 & 1.34 & 30, 44, 70, 100~GHz\\\hline
12 & 4097 & 1.28 & 30, 44, 70, 100~GHz \\\hline
\end{tabular}%
\vfill
\end{minipage}
\hfill
\begin{minipage}[!t]{0.54\textwidth}
\centering
\vspace{14pt}
\includegraphics[width=\textwidth]{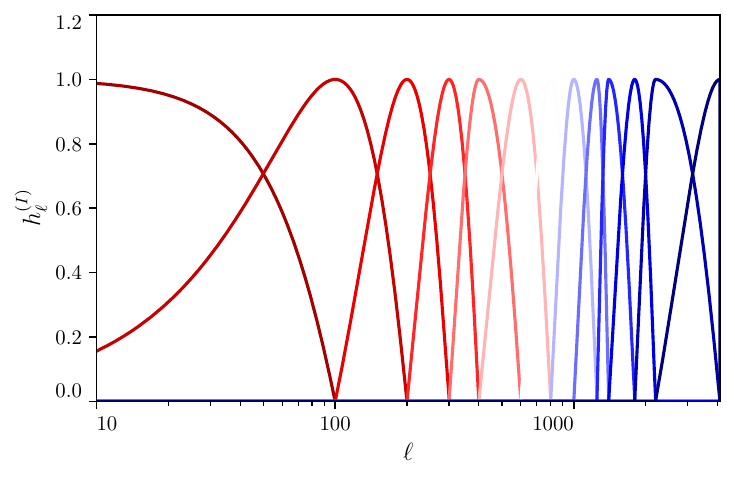}%
\end{minipage}

\caption{
\emph{Left:} properties of the 13 cosine needlets used in our NILC maps, showing each filter's peak multipole, real-space smoothing scale, and excluded \emph{Planck} channels. \emph{Right:} harmonic-space needlet filters $h^{(I)}_\ell$ used in this work.
}
\label{fig:needlet_table_and_filters}

\end{figure*}

\subsection{Inpainting and preprocessing}

\noindent Prior to running the component-separation pipeline, we preprocess the single-frequency NPIPE maps. First, we subtract the Solar dipole and an overall monopole from the maps.\footnote{We use the \texttt{Commander} Solar dipole estimate~\cite{Planck:2020olo}.} Next, we inpaint regions dominated by bright Galactic emission or point sources to prevent them from contributing to the component-separation weights. Following Refs.~\cite{McCarthy:2023hpa, McCarthy:2024ozh, Goldstein:2024mfp}, we apply a diffusive inpainting scheme in which masked pixels are iteratively replaced with the mean of neighboring unmasked pixels. Our inpainting mask, which is shown in the top-left panel of Fig.~\ref{fig:masks_maps_grid}, consists of the union of the \emph{Planck} LFI and HFI point source masks and the Galactic plane mask from the \emph{Planck} 2015 NILC tSZ analysis~\cite{Planck:2015vgm}. After inpainting, we reconvolve all maps to a common Gaussian beam with a full-width half maximum (FWHM) of 5 arcminutes.\footnote{We approximate the \emph{Planck} instrument beams~\cite{Planck:2015wtm, Planck:2015aiq} as Gaussians, with FWHM values given in Table I of Ref.~\cite{McCarthy:2023hpa}.}

\subsection{Component separation}

\noindent Because the kSZ effect preserves the blackbody spectrum of the CMB, it can be isolated from many other microwave sky signals, such as the thermal Sunyaev-Zel'dovich (tSZ) effect, via multi-frequency component separation. Here, we build blackbody temperature maps using the internal linear combination (ILC) method~\cite{COBE_1992, WMAP:2003cmr, Tegmark:2003ve, Eriksen:2004jg}, which constructs the minimum-variance linear combination of the frequency maps that preserves the spectral energy distribution (SED) of the target signal. Since we stack on clusters with large tSZ contributions, we use a constrained ILC to explicitly deproject the tSZ SED. The resulting map has no response to the tSZ SED, albeit at the cost of increased noise. We discuss the impact of tSZ deprojection in Sec.~\ref{sec:tSZ}.

An important element in the ILC approach is choosing the basis in which the covariance is computed. Here, we use needlet ILC (NILC)~\cite{Delabrouille:2008qd}, which computes the weights on a frame of needlets. Needlets are spherical wavelets with compact support in both real and harmonic space~\cite{doi:10.1137/040614359}, allowing the weights to vary with both position and angular scale. Thus, NILC is particularly effective for CMB analyses, where Galactic contributions are often localized in real space and extragalactic contributions are better modeled in harmonic space. 

We use cosine needlets, which can be written in harmonic space as $h^{(I)}_\ell:$
\begin{equation}
h^{(I)}_\ell= \begin{cases}
\cos\left(\frac{\pi}{2}\frac{\ell^I_{\rm peak}-\ell}{\ell^I_{\rm peak}-\ell^{I-1}_{\rm peak}}\right)&\ell^{I-1}_{\rm peak}\le\ell<\ell^{I}_{\rm peak}\\
\cos\left(\frac{\pi}{2}\frac{\ell-\ell^I_{\rm peak}}{\ell^{I+1}_{\rm peak}-\ell^I_{\rm peak}}\right)&\ell^{I}_{\rm peak}\le\ell<\ell^{I+1}_{\rm peak}\\
0 & {\rm otherwise},
\end{cases}
\end{equation}
where $I$ denotes the needlet scale. Following Refs.~\cite{McCarthy:2024ozh, Goldstein:2024mfp}, we use 13 cosine needlets with $\ell_{\rm peak}$ values between $0$ and $4097.$ We perform the component separation procedure on the \emph{Planck} 30, 44, 70, 100, 143, 217, 353, and 545 GHz maps.\footnote{We compute the frequency responses for all components using the full \emph{Planck} passbands as described in Refs.~\cite{McCarthy:2023hpa,McCarthy:2024ozh}.}. We exclude channels whose beam sizes are much larger than the angular scales probed by a given needlet filter, as listed in Table~\ref{fig:needlet_table_and_filters}, following the criterion in Ref.~\cite{McCarthy:2023hpa}. Finally, ILC-based methods are known to have a residual bias because the frequency-frequency covariance is estimated directly from the data (see Ref.~\cite{McCarthy:2023hpa} for a review)~\cite{Delabrouille:2008qd}. This ``ILC bias" can be largely mitigated by choosing a sufficiently large real-space domain on which to infer the ILC weights from the needlet coefficient maps. Here, following Refs.~\cite{McCarthy:2023hpa, McCarthy:2024ozh, Goldstein:2024mfp}, we smooth the needlet coefficient maps with a real-space Gaussian kernel at each scale that ensures the ILC bias is less than 1\% in our final maps.  The left panel of Fig.~\ref{fig:needlet_table_and_filters} summarizes the needlets used in this work, and the right panel shows the harmonic-space needlet filters.

\begin{figure}[!t]
\centering
\includegraphics[width=0.99\linewidth]{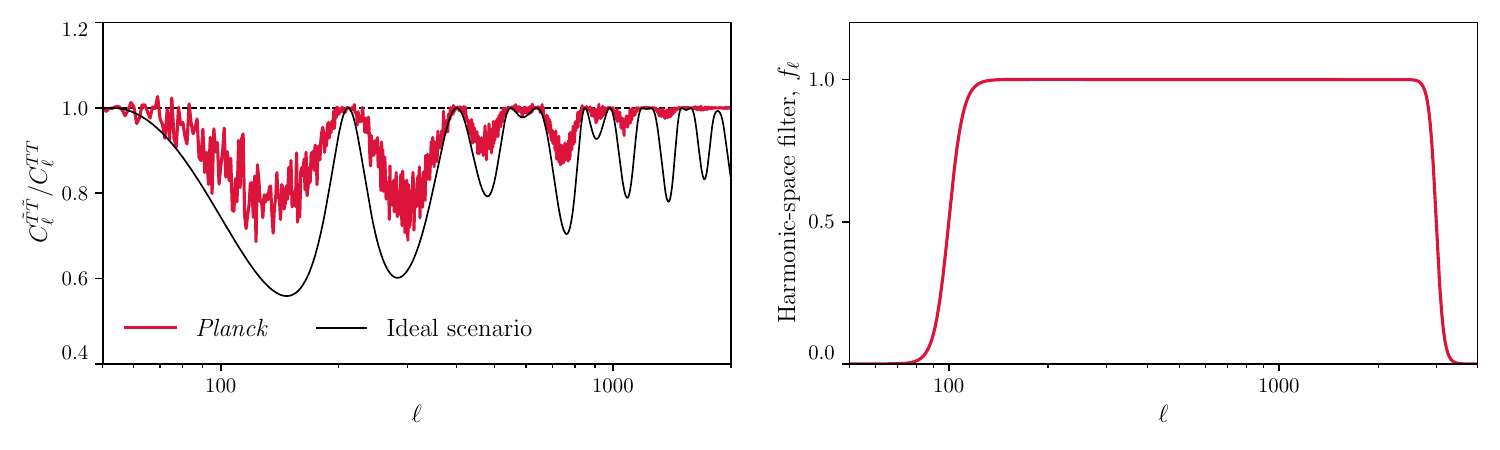}
\caption{\emph{Left:} ratio of the power spectrum of the E-mode-subtracted NILC blackbody temperature map, $\tilde{T}$, to that of the raw NILC $T$ map. A cosmic-variance-limited E-mode survey could achieve significantly better variance reduction (black curve). \emph{Right:} harmonic-space filter applied to remove large- and small-scale modes prior to oriented stacking.}
\label{fig:variance_reduction}
\end{figure}

\subsection{E-mode subtraction and harmonic-space filtering}

\noindent Having constructed the component-separated blackbody temperature map, we apply several filtering steps to suppress the contributions from the primary CMB and noise, while preserving the rkSZ signal. In our fiducial analysis, we build an ``E-mode-subtracted" temperature map, $\tilde{T}$, defined in harmonic space as
\begin{equation}\label{eq:Emode_sub_SM}
    a_{\ell m}^{\tilde{T}}=a_{\ell m}^{T}-\frac{{C}_{\ell}^{TE;~{\rm obs}}}{C_{\ell}^{EE;~{\rm obs}}}\;a_{\ell m}^{E},
\end{equation}
where $a_{\ell m}^T$ and $a_{\ell m}^E$ are respectively the spherical harmonic coefficients of the tSZ-deprojected blackbody temperature and polarization maps, and $C_{\ell}^{EE;~{\rm obs}}$ and $C_{\ell}^{TE;~{\rm obs}}$ are the \emph{observed} E-mode auto-spectrum and temperature-E-mode cross-power spectrum, respectively. 

\begin{figure}[!t]
\centering

\includegraphics[width=0.99\linewidth]{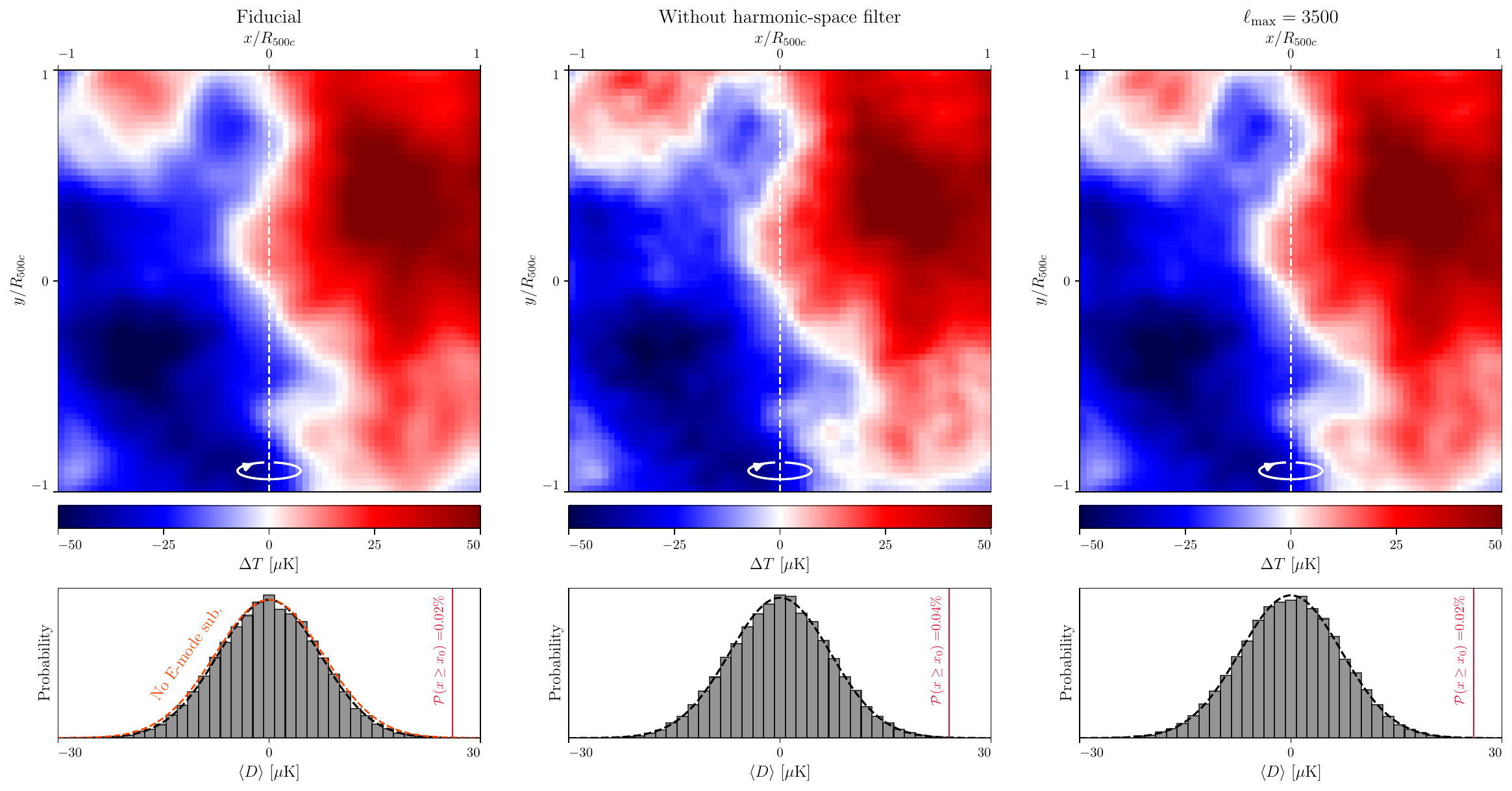}\\[0.6em]
\includegraphics[width=0.99\linewidth]{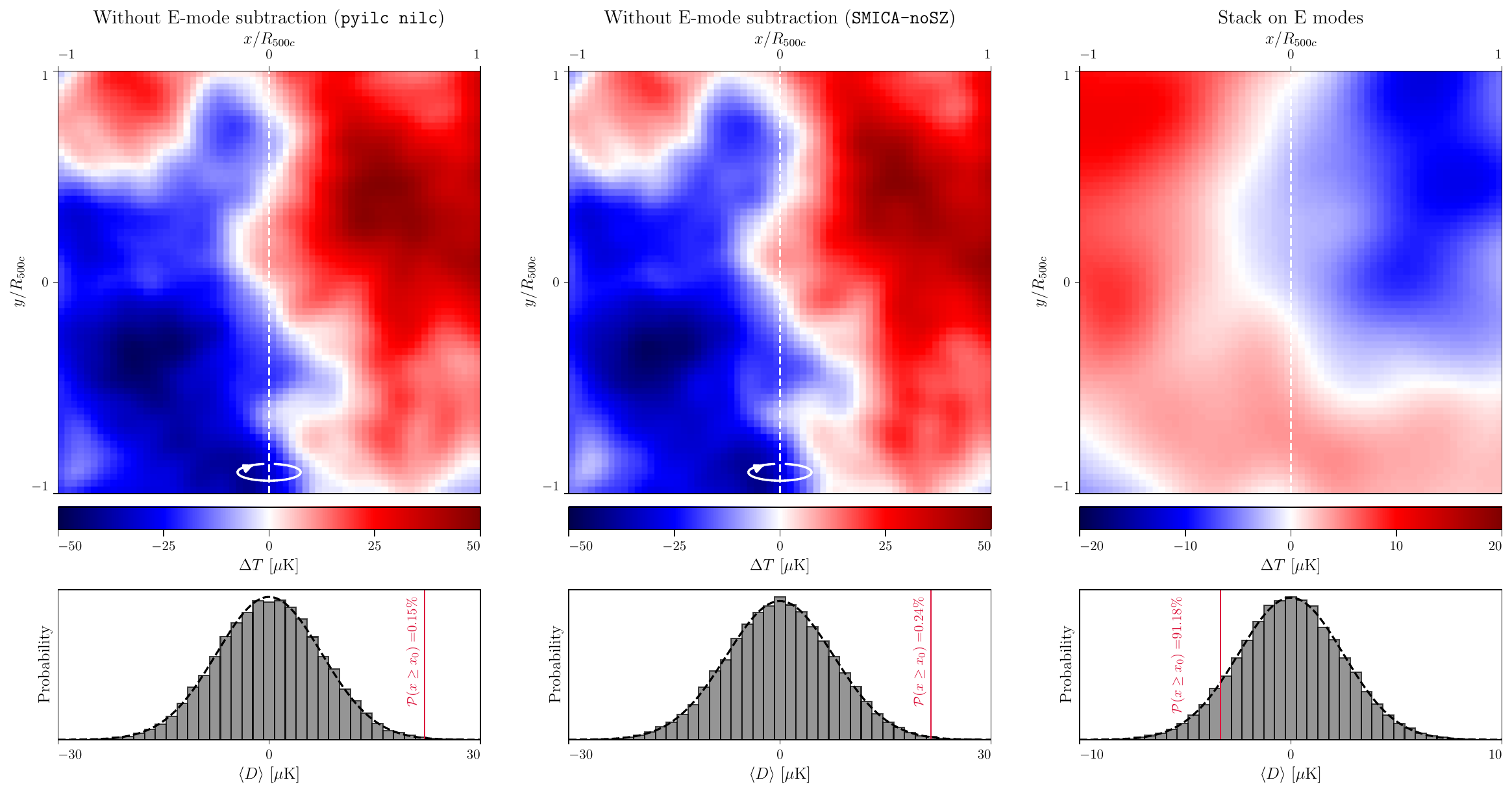}

\caption{Comparison of the estimated dipole for different blackbody temperature maps. The top-left panel shows our fiducial analysis with $3.6\sigma$ evidence for a dipole. The orange-dashed curve shows the Gaussian fit to the distribution of randoms using the temperature map \emph{without} E-mode subtraction, which has $\sim 6\%$ larger noise than our E-mode subtracted map. The top-middle panel illustrates the impact of the harmonic-space filter (Eq.~\eqref{eq:harmonic_space_filter}). Filtering the maps leads to a mild ($0.2\sigma$) increase in the significance. The top-right panel illustrates that extending the harmonic-space filter to $\ell_{\rm max}=3500$, as opposed to $\ell_{\rm max}=3000$, has no impact on the results. The bottom panels illustrate the impact of subtracting E-modes. Stacking on a map without E-mode subtraction reduces the significance from $3.6\sigma$ to $3.0\sigma$. The improvement due to the E-mode subtraction arises from two effects: (i) the lower noise in the E-mode subtracted maps and (ii) the presence of an insignificant ($1.4\sigma$) dipole in the E-mode map that happens to lie in the opposite direction of the expected rkSZ signal, as shown in the bottom-right panel. Finally, the bottom middle panel shows a stack on the \texttt{SMICA-noSZ} map instead of our fiducial \texttt{pyilc} NILC map. The results are consistent with the \texttt{pyilc} analysis, albeit with slightly lower ($0.2\sigma$) significance due to the higher noise in \emph{Planck} PR3 compared to PR4.}

\label{fig:vary_temp_map}
\end{figure}

In Fig.~\ref{fig:variance_reduction}, we show the ratio of the power spectrum of the E-mode-subtracted map to that of the original NILC temperature map. Using \emph{Planck} E-mode data, we find a 10-20\% reduction in the variance in our temperature maps due to primary CMB fluctuations on scales $100\lesssim \ell \lesssim 1000$, which are particularly useful for isolating the kSZ signal in nearby clusters. Since the \emph{Planck} E-mode observations are noise-dominated across all scales, this E-mode cleaning procedure will be even more effective with high-resolution, low-noise E-mode measurements from experiments such as the Atacama Cosmology Telescope~\cite{ACT:2023wcq, AtacamaCosmologyTelescope:2025vnj}, the South Pole Telescope~\cite{SPT-3G:2025bzu}, and the Simons Observatory~\cite{SimonsObservatory:2018koc,SimonsObservatory:2025wwn}. Indeed, the overall variance reduction is set by the cross-correlation coefficient between the temperature and E-mode polarization maps, $r_{\ell}^{TE}\equiv C_{\ell}^{TE}/\sqrt{C_{\ell}^{TT}C_{\ell}^{EE}}$, yielding
\begin{equation}
    C_{\ell}^{\tilde{T}\tilde{T}}=\left( 1-(r_{\ell}^{TE})^2 \right)C_{\ell}^{TT}.
\end{equation}
The black curve in Fig.~\ref{fig:variance_reduction} illustrates the potential variance reduction using signal-dominated E-mode measurements. In principle, $r_{\ell}^{TE}$ can be as high as 40\%, enabling a significant reduction in the primary CMB contributions to the blackbody temperature map, which could be useful in many secondary anisotropy analyses.

In addition to removing the correlated E-mode contribution from our maps, we also apply a harmonic-space filter to suppress large-scale primary CMB fluctuations and small-scale noise in the component-separated map. To this end, we use the following filter
\begin{equation}\label{eq:harmonic_space_filter}
    f_{\ell}=\left[1-\frac{1}{1+\left(\frac{\ell}{\ell_{\rm min}}\right)^{20}} \right]\times \frac{1}{1+\left(\frac{\ell}{\ell_{\rm max}}\right)^{40}}
\end{equation}
with $\ell_{\rm min}=100$ and $\ell_{\rm max}=3000.$ We show the harmonic-space filter in the right panel of Fig.~\ref{fig:variance_reduction}

In Fig.~\ref{fig:vary_temp_map}, we assess the impact of the harmonic-space filter and E-mode subtraction on the significance of our rkSZ dipole measurement. Without the harmonic-space filter (top middle), the significance decreases from $3.6\sigma$ to $3.4\sigma$. Furthermore, our detection significance is largely insensitive to the maximum multipole in the filter, as we find the same level of significance using $\ell_{\rm max}=3500$ (top right) and $\ell_{\rm max}=2500$ (not shown). Ultimately, these tests demonstrate that the dipole in our stack is not an artifact of the filtering procedure, and that our harmonic-space filter has a small, but positive, impact on the signal-to-noise.

The bottom panel of Fig.~\ref{fig:vary_temp_map} quantifies the impact of removing E-modes on the rkSZ dipole detection significance. Without subtracting E-modes,\footnote{All stacks in this row still use the harmonic-space filter.} the significance decreases from $3.6\sigma$ to $3.0\sigma$. To quantify how much of the improvement is coming from the reduced noise in the E-mode subtracted map, the orange distribution in the top-left panel shows the Gaussian fit to the randoms estimated from the temperature map without E-mode subtraction.\footnote{Recall that our optimal weights depend on the particular map via the $\sigma_{\tilde{T}}^2(M_{\rm 500c}, z)$ contribution in Eq.~\eqref{eq:fid_weights}. Thus, to isolate the improvement from using $\tilde{T}$ instead of ${T}$ due solely to the reduction in noise, we plot the distribution using the optimal weights derived from the $\tilde{T}$ analysis. Consequently, the orange curve in the top-left panel is \emph{slightly} different from the black curve in the bottom-left panel, which uses weights that are optimized for the $T$ map. } The distribution is $\sim6\%$ broader than the E-mode-subtracted map, which corresponds to $\approx 0.2-0.3\sigma$ decrease in the dipole significance. 

To understand the remaining improvement, we perform an oriented and weighted stack on the filtered E-mode map, $a_{\ell m}^{\tilde{E}}=f_{\ell}\frac{{C}_{\ell}^{TE;~{\rm obs}}}{C_{\ell}^{EE;~{\rm obs}}}\;a_{\ell m}^{E}$, using the fiducial weights. In principle, this should be uncorrelated with the cluster rotation. However, there is a mild dipole (bottom-right panel) in the filtered E-mode map in the opposite direction of the expected rkSZ dipole. This dipole, which is statistically insignificant ($1.4\sigma)$, explains why the E-mode subtraction improves the detection significance slightly beyond the simple variance-based expectation, i.e., this is due to a fortunate fluctuation. Ultimately, it is reassuring that a significant dipole persists in the stack without E-mode subtraction and harmonic-space filtering, and encouraging that the signal-to-noise of the dipole \emph{increases} after we apply these physically-motivated procedures designed to extract the rkSZ signal.

Finally, the bottom-middle panel of Fig.~\ref{fig:vary_temp_map} shows the stack on the \texttt{SMICA-noSZ} map, as was used in Ref.~\cite{Baxter:2019tze}. The results are consistent with those from the NILC maps used in this work. The dipole in \texttt{SMICA-noSZ} is slightly ($\approx 0.2\sigma$) less significant than in our \texttt{pyilc} stacks, which is consistent with the improvements expected from using \emph{Planck} PR4 instead of PR3 data.

\clearpage
\pagebreak

\begin{figure}[!t]
\centering
\includegraphics[width=0.99\linewidth]{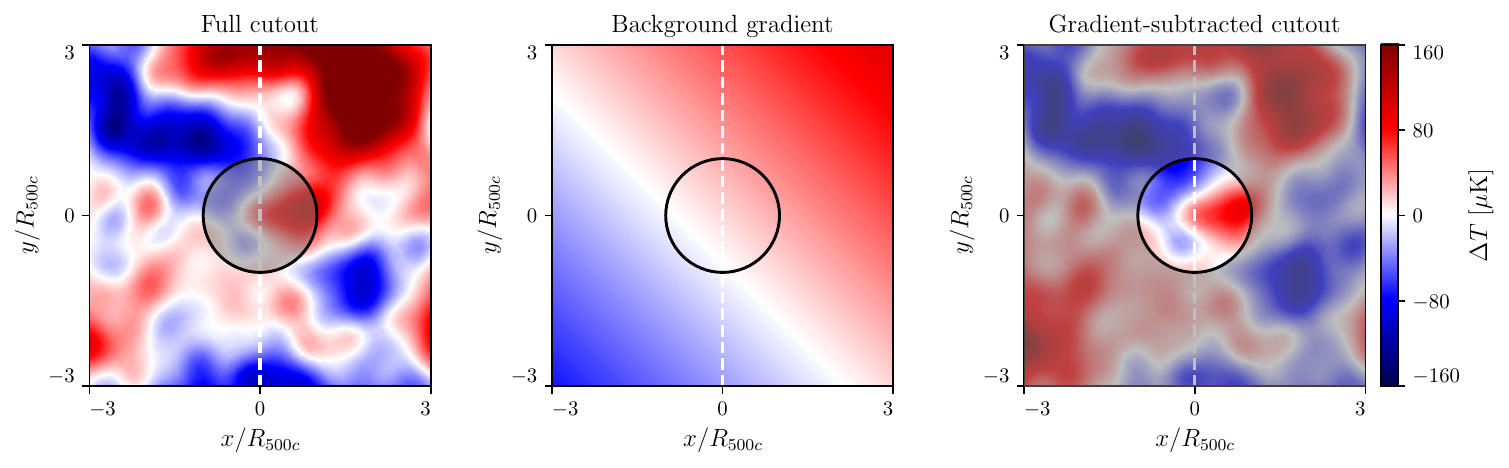}
\caption{Example of our gradient-subtraction procedure applied to A1750. The left panel shows an extended cutout around the cluster, including pixels out to $3R_{\rm 500c}$. The shaded region denotes $R \leq R_{\rm 500c}$, which is excluded from the gradient estimation. The middle panel shows the best-fit background-gradient estimate, while the right panel shows the cutout after subtracting this estimate. For clarity, we have shaded $R\geq R_{\rm 500c}$ in the right panel. Notice that the background subtraction significantly reduces the large-scale gradient across the entire cutout.}
\label{fig:background_gradient_subtract_example}
\end{figure}
\begin{figure}[!t]
\centering
\includegraphics[width=0.67\linewidth]{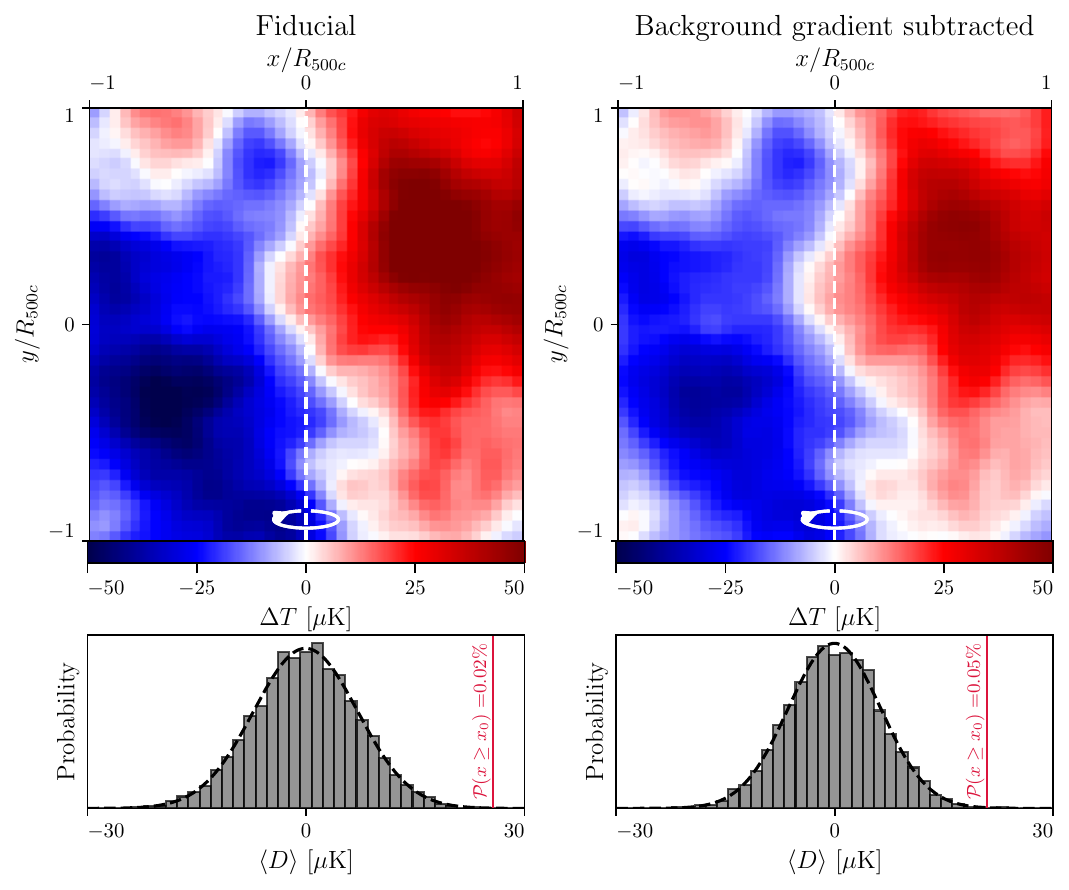}
\caption{Comparison of stack before (left) and after (right) applying the background-gradient subtraction method outlined in Sec.~\ref{App:grad_background_sub}. After subtracting the estimated background gradients from each cutout, the estimated dipole amplitude decreases by approximately 20\% and the noise decreases by 10\%. Consequently, the resulting dipole remains statistically significant at $3.3\sigma$.}
\label{fig:background_gradient_subtraction_stack}
\end{figure}
%

\subsection{Robustness to the primary CMB and foregrounds}

%
\subsubsection{Primary CMB}\label{App:grad_background_sub}

\noindent The dominant source of noise in our measurement comes from the primary CMB, which is approximately a gradient over the range of scales considered in this work. In this section, we present a method to estimate and mitigate the primary CMB gradient inspired by aperture photometry (see, \emph{e.g.}, Ref.~\cite{AtacamaCosmologyTelescope:2020wtv}). We stress that the primary CMB contributions are already included in our noise model. Furthermore, the primary CMB is naturally down weighted relative to the kSZ signal via our harmonic-space filtering, $E$-mode subtraction, and redshift-dependent weights. Nevertheless, we perform this further test to estimate how much of our signal could be due to random large-scale CMB fluctuations that extend beyond the cluster scale.

We start by extracting extended cutouts of the clusters including pixels within a square region with a half-width of $3R_{\rm 500c}$. In order to estimate the background gradient, we fit a three-parameter linear model $(\Delta \tilde{T}(x,y)=ax+by+c)$ to all pixels with $R\geq R_{\rm 500c}$\footnote{We exclude pixels with $R < R_{\rm 500c}$ from the fit to reduce contamination of the background estimate by the true cluster signal. We have tested larger exclusion radii and find that this choice has a negligible impact on the final results.} using ordinary least-squares regression. Finally, we subtract the background estimate from the cutout to obtain a ``gradient-subtracted cutout." Fig.~\ref{fig:background_gradient_subtract_example} shows an example of this procedure for the cluster with the second highest weight in our sample, A1750.\footnote{We do not show the background subtraction for the highest-weight cluster in our sample, Coma, because it spans such a large angular scale that the background CMB cannot be accurately modeled via a simple gradient. In the future, it would be interesting to explore more sophisticated background-subtraction methods that explicitly account for the angular scales of each cutout.}

We apply the background-subtraction method to all cluster cutouts, including the randoms,\footnote{We use slightly lower-resolution cutouts than in the main text and estimate the noise using 5,000 random cutouts because it is computationally expensive to extract the larger patches required in this section. We have checked that these choices have no impact on our final results.} and repeat the stack. Note that since our weights include a Monte-Carlo contribution based on the likelihood of finding a dipole in a cutout of a particular angular scale, which is set by the cluster redshift and $R_{\rm 500c}$, the optimal weights for the background-subtracted cutouts are different from those used in the fiducial analysis. Therefore, we recompute the optimal weights for the background-subtracted cutouts, although we find that this does not have an appreciable impact on our results.

Fig.~\ref{fig:background_gradient_subtraction_stack} illustrates the impact of the background-gradient subtraction method on our estimated dipole amplitude. Although the background-gradient subtraction reduces the amplitude of the dipole by approximately $20\%$, the noise distribution is also reduced by $10\%$, hence the dipole signal discussed in the main text persists at $3.3\sigma$. This test further demonstrates that the signal cannot be solely attributed to chance fluctuations of large-scale primary CMB gradients.

\begin{figure}[!t]
\centering
\includegraphics[width=0.99\linewidth]{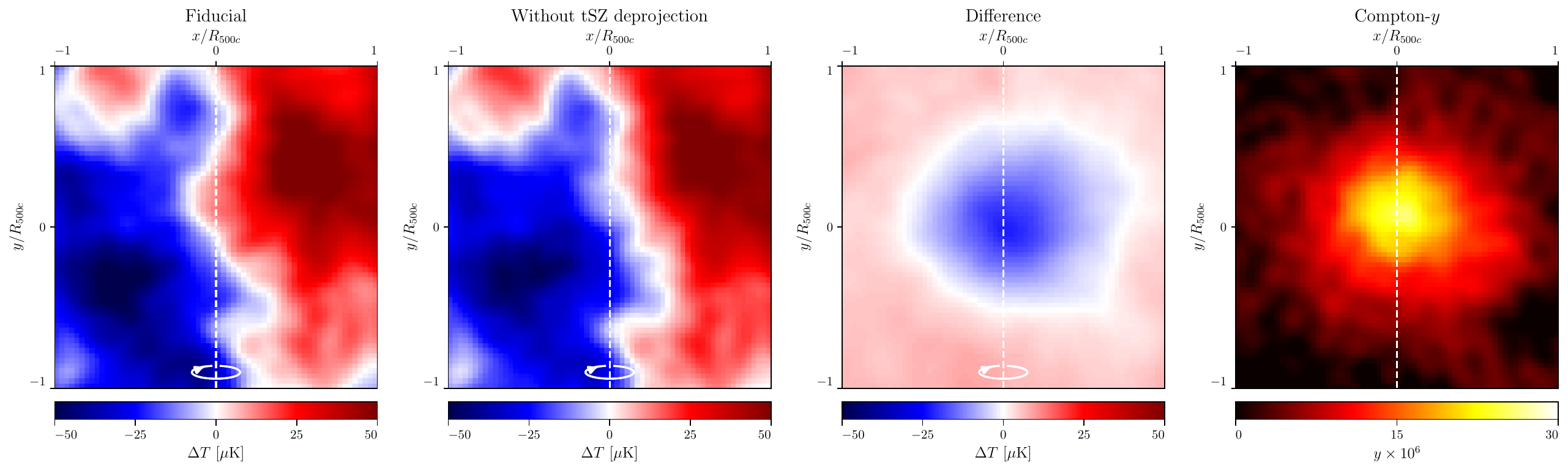}
\caption{Impact of the thermal Sunyaev–Zel’dovich (tSZ) effect on the oriented stack. The first panel shows our fiducial stack, which explicitly deprojects the tSZ contribution to the NILC temperature map. The second panel shows the same analysis applied to a map without tSZ deprojection. While both stacks display a clear dipole, the map without tSZ deprojection contains residual tSZ contributions, which can be seen clearly by taking the difference between the two stacks (third panel). For reference, the final panel shows the oriented stack measured on a NILC Compton-$y$ (tSZ) map, in which no dipole is seen.}
\label{fig:tSZ_stacks}
\end{figure}
\subsubsection{tSZ effect}\label{sec:tSZ}

\noindent Since we stack on massive clusters, the single-frequency maps entering our component-separation pipeline could contain large tSZ contributions. Although the tSZ signal is not dipolar and therefore should not bias our dipole estimate directly, it does generate additional noise at the cluster locations. To assess the impact of tSZ contamination, we repeat our stacking analysis on (i) component-separated maps constructed without tSZ deprojection and (ii) component-separated Compton-$y$ maps. These results are shown in Fig.~\ref{fig:tSZ_stacks}. While the dipole is evident in both our tSZ-deprojected and non-tSZ-deprojected stacks,\footnote{The non-tSZ-deprojected map has a dipole significance of $3.4\sigma$ based on the randoms procedure used in this work. However, we note that a rigorous assessment of the dipole significance in this map requires a more careful noise treatment, because the residual tSZ contributions introduce cluster-correlated noise.} the map without tSZ deprojection shows clear residual tSZ leakage. This is most apparent in the difference between the two stacks (third panel), which reveals a prominent cold spot originating from the tSZ signal in the single-frequency NPIPE maps below 217~GHz. The shape of this residual closely matches the stack on the Compton-$y$ map shown in the right panel. The fact that both the residual and Compton-$y$ stacks are well-centered further confirms that our X-ray cross-matching procedure accurately identifies the cluster centers, thus demonstrating that our dipolar signal is not an artifact of miscentering.

\begin{figure}[!t]
\centering
\includegraphics[width=0.99\linewidth]{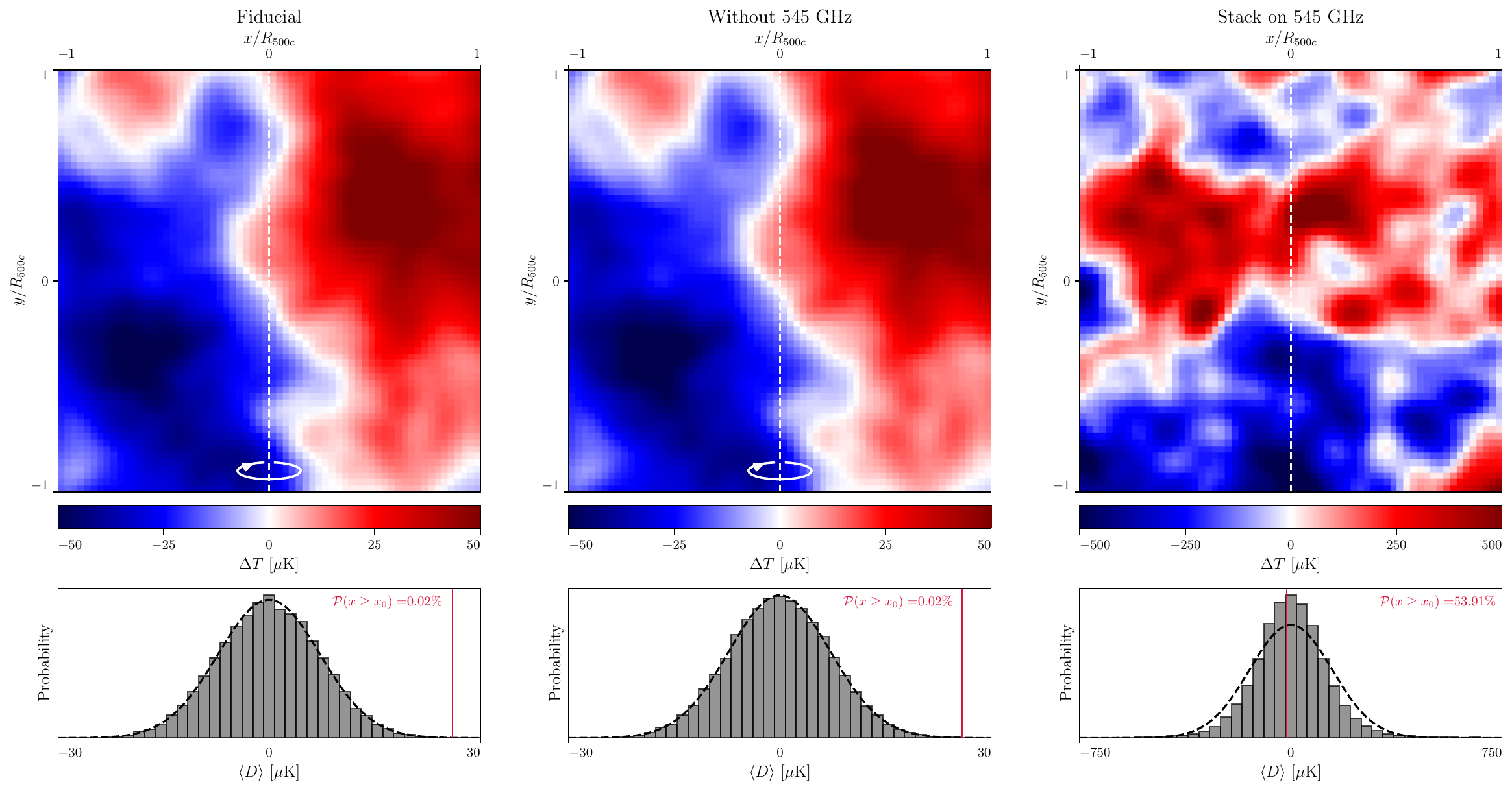}
\caption{Impact of dust emission on the oriented stack. The first panel shows our fiducial stack, which includes the \emph{Planck} 545 GHz channel. The middle panel shows the same analysis applied to a NILC map that does not include the \emph{Planck} 545 GHz data. Including the 545 GHz channel has a negligible effect on both the stack and the dipole amplitude significance. The right panel shows an oriented stack on the \emph{Planck} 545 GHz map. We find no evidence of a rotation-aligned dipole in the 545 GHz map. Note that the randoms distribution for the 545 GHz map is non-Gaussian due to the non-Gaussianity of the dust.}
\label{fig:dust_stacks}
\end{figure}
%

\subsubsection{Doppler-boosted dust emission}

\noindent The Doppler boosting of dust emission from galaxies in clusters can also produce a dipole that is correlated with cluster rotation~\cite{Maniyar:2022lkv}. This effect is expected to be small for the low-redshift, massive cluster sample used here, as the galaxies in such clusters are typically ``red and dead'', with very little star formation and thermal dust emission~\cite[e.g.,][]{1992MNRAS.254..601B,1996MNRAS.283.1361B,2006MNRAS.373..469B}. Nevertheless, we conduct two tests to explicitly demonstrate that the dipole in our oriented stacks is not caused by residual dust emission in the NILC temperature map. First, we repeat our stack using a component-separated map that excludes the dust-dominated 545 GHz data. The stack is shown in the middle panel of Fig.~\ref{fig:dust_stacks}. Excluding the 545~GHz map has no impact on the resulting dipole significance. Second, we repeat the oriented stacking pipeline directly on the harmonic-space-filtered 545~GHz map, using the weights from the fiducial stack. We find no dipole signal in the stacked 545 GHz map. Note that, in this case, the distribution of randoms is not well described by a Gaussian because of the intrinsic non-Gaussianity of the dust distribution in the maps. In conclusion, these tests demonstrate that our dipole signal cannot be attributed to Doppler boosting of dust emission.

\clearpage
\pagebreak

\begin{figure}[!t]
\centering
\includegraphics[width=0.99\linewidth]{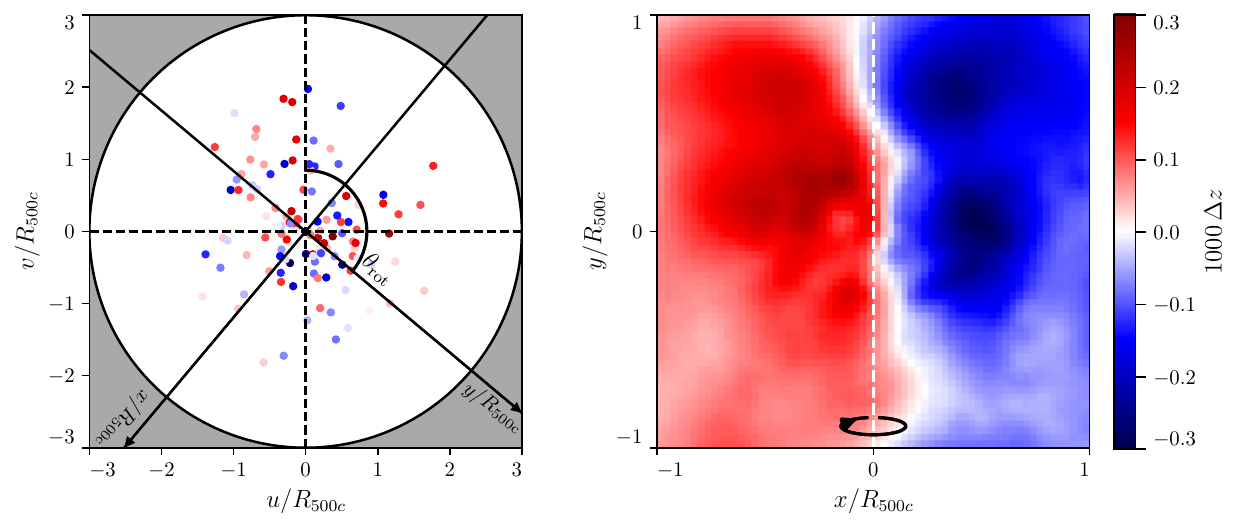}
\caption{\emph{Left:} geometry and variables used to estimate the rotation direction of the J1217.6+0339 cluster. We project member galaxies from the group catalog onto the tangent plane centered on the cluster's X-ray position. Here, $u$ and $v$ represent the coordinates in the tangent plane. The points mark the projected galaxy positions, with colors indicating each galaxy's redshift relative to the group mean. The estimated rotation direction $\theta_{\rm rot}$ is defined as the angle that maximizes the difference in mean redshift between galaxies split into two hemispheres. The oriented ($x-y$) coordinate system is defined such that the $y$-axis is aligned with $\theta_{\rm rot}$ with galaxies to the left (right) of the $y$-axis having, on average, a redshift (blueshift) relative to the mean redshift of all galaxies within the cluster. We exclude galaxies beyond $3\,R_{\rm 500c}$, indicated by the shaded gray region. \emph{Right: } oriented stack of the member-galaxy redshift map. We recover a significant dipole. Note that we convolve the galaxy redshift map with a 5 arcmin beam to match the resolution of the \emph{Planck} data.}
\label{fig:coord_system_def}
\end{figure}
%

\section{Additional details on cluster rotation estimate, stacking pipeline, and cluster sample}

\subsection{Cluster rotation direction estimate and stacking pipeline}\label{subsec:angle_ident}

\noindent We estimate each cluster's rotation direction based on the method from Ref.~\cite{Tang:2025lmm}. The left panel of Fig.~\ref{fig:coord_system_def} illustrates the coordinate system and angle-identification procedure used for an example cluster, J1217.6+0339. For each cluster, we project all member galaxies onto the tangent plane centered on the X-ray center (i.e., the location of peak X-ray luminosity) using the \texttt{Coord} package.\footnote{\href{https://github.com/LSSTDESC/Coord}{https://github.com/LSSTDESC/Coord}} We compute the projected distance between each galaxy and the cluster center using the angular diameter distance to the cluster redshift,\footnote{Aside from using this distance to exclude a very small set of galaxies beyond 3\,$R_{\rm 500c}$ from our rotation estimates, the entire angle identification procedure is independent of the distance to the cluster.} obtained from the X-ray catalog. We select all galaxies within a projected distance of $3\,R_{\rm 500c}$ to estimate the rotation direction.

Within the tangent plane, we define coordinates $u$ and $v$ such that the $u$-axis points toward decreasing right ascension (west) and the $v$-axis toward increasing declination (north). We split the galaxy sample into two hemispheres separated by a ray passing through the origin with an angle $\theta_{\rm trial}$ defined clockwise with respect to the $v$-axis and compute the mean redshift in each hemisphere. We compute the difference between the left and right hemispheres and repeat this procedure for $0^\circ \leq \theta_{\rm trial} < 360^\circ$ in steps of $5^\circ$. We define $\theta_{\rm rot}$ as the angle that maximizes this difference, which we denote by $\Delta z_{\rm dip}^{\rm max}$. Having estimated the cluster's rotation direction, we define an oriented coordinate system in which the $y$-axis is aligned with the estimated rotation direction. On average, galaxies to the left (right) of the $y$-axis have a positive (negative) redshift compared to the mean redshift of all member galaxies within the cluster.

To validate our angle-identification and oriented stacking pipeline, it is useful look at the full sample of 25 clusters. To this end, we construct a HEALPix map of the member-galaxy redshifts for the cluster sample. Starting from an empty map, we assign each pixel corresponding to a member galaxy with a value equal to the difference between that galaxy's redshift and the mean redshift of all member galaxies in its group. We convolve this map with a 5 arcmin Gaussian beam to match the resolution of the \emph{Planck} data. The right panel of Fig.~\ref{fig:coord_system_def} shows the oriented stack applied to this redshift map.\footnote{We do not apply the rkSZ-optimized weights here because the goal of this test is simply to validate the angle-identification and oriented stacking pipeline.} As expected, we find a significant dipole aligned with the estimated rotation axis. Note that the sign of the dipole is the opposite of that of the rkSZ dipole, since a receding galaxy corresponds to a positive redshift but a negative rkSZ signal.

\begin{figure}[!t]
\centering
\includegraphics[width=0.99\linewidth]{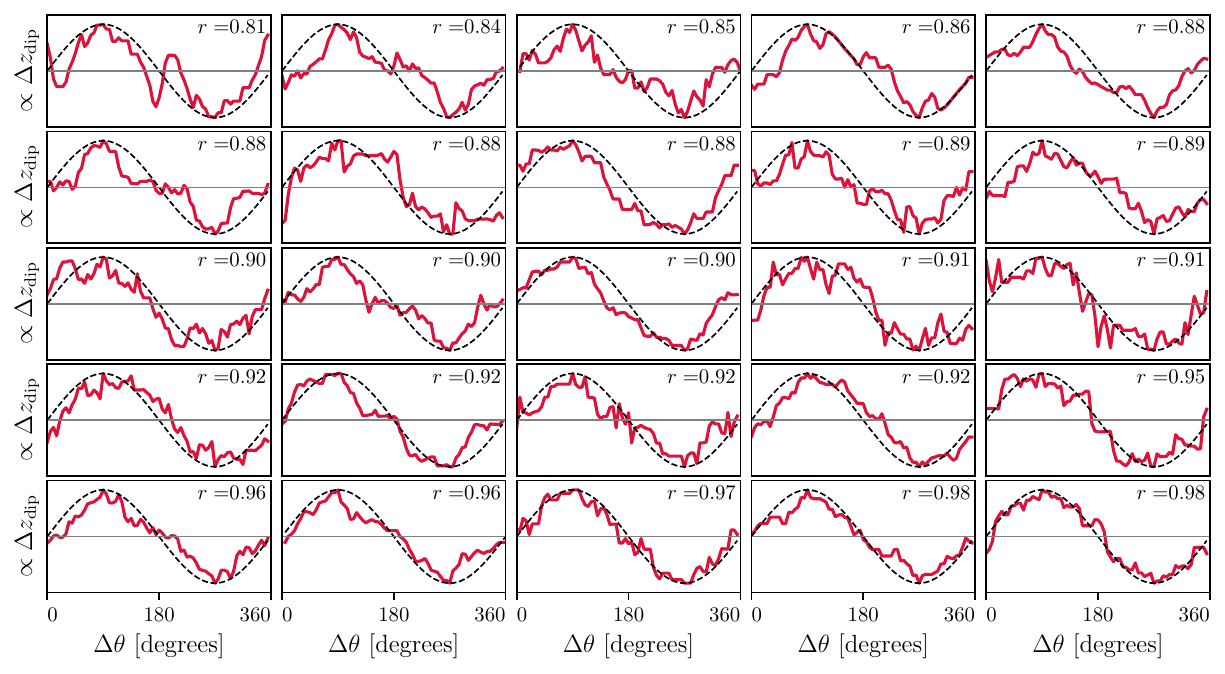}
\caption{Difference in the average redshift between the left and right hemispheres as a function of the trial rotation angle $\Delta \theta$ for our sample of 25 clusters. For visualization, we shift each curve so that the maximum value occurs at $\Delta\theta = 90^\circ$. We compare the measured rotation signal to a sinusoidal template (black dashed) and show the Pearson correlation coefficient between the two in the upper-right corner of each panel. The panels are sorted in order of increasing correlation coefficient.}
\label{fig:rotation_angle_dependence}
\end{figure}

Fig.~\ref{fig:rotation_angle_dependence} shows the dipole amplitude as a function of the trial rotation angle for our 25 clusters. We shift each curve such that the maximum value occurs at $\Delta \theta=90^\circ$. The black dashed line shows a sinusoidal template with amplitude $\Delta z_{\rm dip}^{\rm max}$ and zero phase. For a purely rotating cluster, $\Delta z_{\rm dip}$ is expected to follow this sinusoidal template. Therefore, we use the Pearson correlation coefficient $r$ between the sinusoidal template and the measured $\Delta z_{\rm dip}(\Delta \theta)$ to estimate the fidelity of the rotation estimate on a per-cluster basis. We discard clusters with $r\leq 0.8.$ This removes clusters where the rotation direction cannot be robustly identified, thus increasing our sensitivity to rkSZ.

As a final validation of our stacking pipeline, we reproduce the results of Ref.~\cite{Baxter:2019tze}. Using the \texttt{SMICA-noSZ} map, the sample of 13 galaxy clusters from Ref.~\cite{Manolopoulou:2016ozk}, and the same weights and pixel-space filter as used in Ref.~\cite{Baxter:2019tze}, we obtain a dipole amplitude of $13.7~\mu{\rm K}$, which exceeds 96.1\% ($\sim 1.8\sigma$) of our randoms. Our results are completely consistent with the quoted dipole amplitude of $27.6~\mu{\rm K}$ with 96\% significance from Ref.~\cite{Baxter:2019tze}, noting that our convention defines the amplitude to be half that used in Ref.~\cite{Baxter:2019tze}.

\clearpage
\pagebreak

\begin{table}\label{table:cluster_properties}
\centering
\setlength{\tabcolsep}{4pt}
\begin{tabular}{|c|c|c|c|c|c|c|c|c|c|c|c|}
\hline
MCXC Name & Alt. Name & RA & Dec & $z$ & $M_{\rm 500c}$ [$M_{\odot}$] & $R_{\rm 500c}$ [Mpc] & $N_{\rm mem}$ & $r$ & $\Delta z_{\rm dip}^{\rm max}$ & $\theta_{\rm rot}$ & $w~[\%]$ \\
\hline
J1132.8+1428 & A1307 & 173.22$^\circ$ & 14.47$^\circ$ & 0.083 & 3.00e+14 & 0.99 & 77 & 0.95 & 1.4e-03 & 40$^\circ$ & 6.5 \\
J1359.2+2758 & A1831 & 209.82$^\circ$ & 27.97$^\circ$ & 0.075 & 2.51e+14 & 0.94 & 68 & 0.88 & 3.3e-03 & 240$^\circ$ & 10.7 \\
J1115.5+5426 &   & 168.89$^\circ$ & 54.44$^\circ$ & 0.069 & 1.07e+14 & 0.71 & 56 & 0.88 & 1.8e-03 & 5$^\circ$ & 2.6 \\
J1241.3+1834 & A1589 & 190.33$^\circ$ & 18.57$^\circ$ & 0.073 & 1.80e+14 & 0.84 & 89 & 0.88 & 2.3e-03 & 340$^\circ$ & 7.4 \\
J1302.8-0230 & A1663 & 195.71$^\circ$ & -2.51$^\circ$ & 0.085 & 1.62e+14 & 0.81 & 73 & 0.92 & 1.8e-03 & 40$^\circ$ & 5.5 \\
J1539.6+2147 & A2107 & 234.91$^\circ$ & 21.79$^\circ$ & 0.041 & 1.51e+14 & 0.80 & 104 & 0.96 & 1.8e-03 & 105$^\circ$ & 3.7 \\
J1529.0+5249 &   & 232.27$^\circ$ & 52.83$^\circ$ & 0.074 & 1.09e+14 & 0.71 & 51 & 0.97 & 1.4e-03 & 55$^\circ$ & 1.9 \\
J1330.8-0152 & A1750 & 202.71$^\circ$ & -1.87$^\circ$ & 0.085 & 3.04e+14 & 1.00 & 94 & 0.86 & 2.4e-03 & 265$^\circ$ & 12.9 \\
J1303.7+1916 & A1668 & 195.94$^\circ$ & 19.27$^\circ$ & 0.064 & 1.89e+14 & 0.86 & 76 & 0.88 & 1.2e-03 & 180$^\circ$ & 2.8 \\
J1512.7+0725 &   & 228.18$^\circ$ & 7.42$^\circ$ & 0.046 & 9.07e+13 & 0.67 & 106 & 0.92 & 7.9e-04 & 50$^\circ$ & 1.3 \\
J1256.9+2724 &   & 194.24$^\circ$ & 27.40$^\circ$ & 0.024 & 4.48e+13 & 0.54 & 229 & 0.84 & 9.2e-04 & 0$^\circ$ & 1.1 \\
J1259.7+2756 & Coma & 194.93$^\circ$ & 27.94$^\circ$ & 0.023 & 4.22e+14 & 1.13 & 223 & 0.92 & 1.8e-03 & 350$^\circ$ & 14.8 \\
J1454.5+1838 & A1991 & 223.63$^\circ$ & 18.64$^\circ$ & 0.059 & 1.68e+14 & 0.82 & 74 & 0.90 & 1.8e-03 & 110$^\circ$ & 3.8 \\
J1426.6+1642 &   & 216.66$^\circ$ & 16.71$^\circ$ & 0.054 & 1.27e+14 & 0.75 & 79 & 0.89 & 8.3e-04 & 195$^\circ$ & 1.4 \\
J1532.4+0446 &   & 233.11$^\circ$ & 4.77$^\circ$ & 0.039 & 1.01e+14 & 0.70 & 58 & 0.98 & 1.0e-03 & 250$^\circ$ & 0.8 \\
J1022.0+3830 & RXJ1022.1+3830 & 155.52$^\circ$ & 38.51$^\circ$ & 0.054 & 8.11e+13 & 0.65 & 67 & 0.85 & 9.3e-04 & 285$^\circ$ & 1.0 \\
J1217.6+0339 & Zw 1215.1+0400 & 184.42$^\circ$ & 3.66$^\circ$ & 0.077 & 3.56e+14 & 1.05 & 119 & 0.90 & 1.5e-03 & 130$^\circ$ & 11.3 \\
J1613.3+3050 &   & 243.33$^\circ$ & 30.84$^\circ$ & 0.052 & 8.08e+13 & 0.65 & 53 & 0.98 & 1.7e-03 & 325$^\circ$ & 1.2 \\
J1523.0+0836 & A2063 & 230.77$^\circ$ & 8.60$^\circ$ & 0.036 & 2.15e+14 & 0.90 & 107 & 0.96 & 1.4e-03 & 130$^\circ$ & 3.6 \\
J1147.3+5544 & A1377 & 176.84$^\circ$ & 55.74$^\circ$ & 0.051 & 7.29e+13 & 0.63 & 80 & 0.91 & 7.1e-04 & 150$^\circ$ & 0.8 \\
J1521.8+0742 & MKW 3s & 230.46$^\circ$ & 7.71$^\circ$ & 0.044 & 2.52e+14 & 0.95 & 85 & 0.90 & 6.9e-04 & 235$^\circ$ & 1.8 \\
J1110.7+2842 & A1185 & 167.70$^\circ$ & 28.71$^\circ$ & 0.031 & 6.21e+13 & 0.60 & 78 & 0.89 & 7.8e-04 & 205$^\circ$ & 0.5 \\
J1116.3+2918 &   & 169.09$^\circ$ & 29.30$^\circ$ & 0.047 & 7.33e+13 & 0.63 & 74 & 0.81 & 1.0e-03 & 30$^\circ$ & 1.0 \\
J1147.9+5441 &   & 176.98$^\circ$ & 54.69$^\circ$ & 0.060 & 1.40e+14 & 0.77 & 68 & 0.92 & 7.5e-04 & 170$^\circ$ & 1.2 \\
J1204.1+2020 & NGC 4066 & 181.05$^\circ$ & 20.35$^\circ$ & 0.025 & 3.25e+13 & 0.48 & 52 & 0.91 & 7.7e-04 & 120$^\circ$ & 0.2 \\
\hline
\end{tabular}
\caption{Properties of the 25 X-ray cross-matched clusters used in this work. The first two columns list the MCXC name and, when available, the alternative from the MCXC-II catalog, respectively. We also include the cluster center, redshift, mass, and radius from the MCXC-II catalog. $N_{\rm mem}$ is the total number of member galaxies used to estimate the rotation direction. $r$ is the Pearson correlation coefficient of the measured cluster rotation estimator with a sinusoidal template (see Sec.~\ref{subsec:angle_ident}). $\Delta z_{\rm dip}^{\rm max}$ is the amplitude of the dipole in the member-galaxy redshift distribution, which serves as an estimate of the rotational velocity. $\theta_{\rm rot}$ is the estimated rotation angle (see Sec.~\ref{subsec:angle_ident} for details regarding the coordinate system). The final column gives the weight assigned to each cluster in our fiducial stack, as computed with Eq.~\eqref{eq:fid_weights_SM}.}
\end{table}

\subsection{Table of cluster properties}

\noindent Table~\ref{table:cluster_properties} lists the properties of the cluster sample used in this work. Note that several of these clusters are known to be undergoing mergers (e.g., Abell 1750)~\cite{Belsole:2005zc}, which could enhance their associated rkSZ signal~\cite{Altamura:2023hoe}.  Due to the moderate significance of our rkSZ dipole detection, we do not attempt to sub-divide the sample into clusters of differing dynamical states, but note that this would be an interesting avenue for follow-up work.

\clearpage
\pagebreak

\subsection{Varying cluster selection}

\begin{figure}[!t]
\centering
\includegraphics[width=0.99\linewidth]{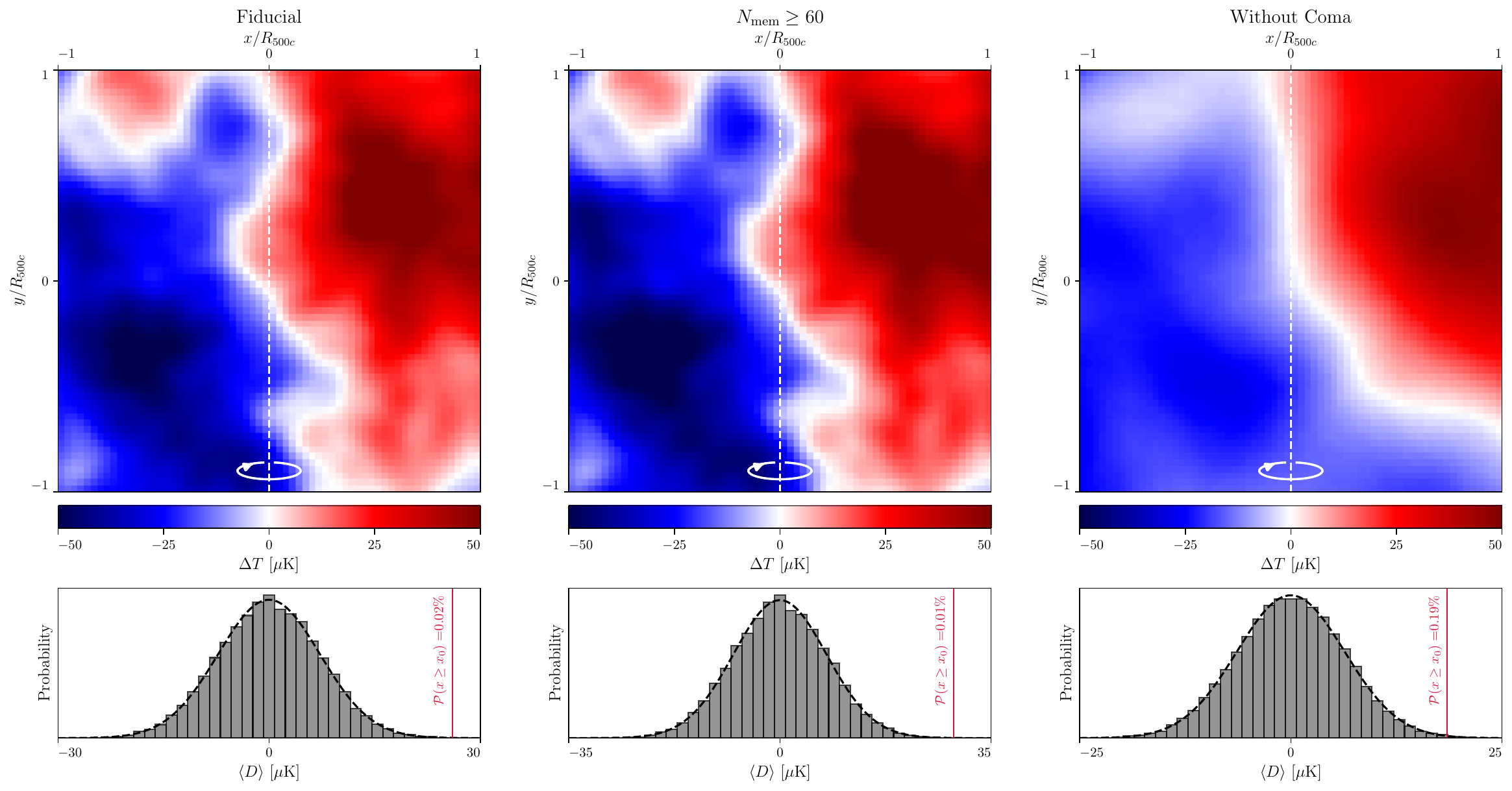}
\caption{Impact of varying the cluster sample used in the oriented stack. The left panel shows our fiducial stack based on 25 clusters. The middle panel excludes five clusters that have fewer than 60 member galaxies. Using this reduced sample leads to a negligible ($\sim 0.1\sigma$) increase in the dipole significance. The right panel removes the Coma cluster, which carries 15\% of the weight in our fiducial stack. Even without Coma, there is still significant ($2.9\sigma$) evidence for a dipole.}
\label{fig:change_cluster_sample}
\end{figure}

\noindent In Fig.~\ref{fig:change_cluster_sample}, we show several oriented stacks using different subsets of our fiducial cluster sample. The middle panel uses clusters with at least $N_{\rm mem}\geq 60$, as opposed to our fiducial analysis with $N_{\rm mem}\geq 50.$ Restricting to this subset of 20 clusters yields a negligible ($\sim 0.1\sigma$) increase in the significance of the inferred dipole. In the right panel, we remove the Coma cluster from our analysis. In our fiducial analysis, Coma carries 15\% of the weight because of its large mass and high number of member galaxies. Thus, we expect Coma to provide a sizable contribution to our estimated dipole. Nevertheless, after removing Coma, we still find significant ($ 2.9\sigma$) evidence for a dipole. Furthermore, the stacks with and without Coma are visually broadly consistent, although the stack without Coma contains fewer small-scale features than our fiducial stack. This arises from two effects: (i) the large weight of a single object will amplify the features associated with that particular object, and, more importantly, (ii) the Coma cutout spans nearly twice the angular size of the other clusters in our sample due to its high mass and low redshift.

\begin{figure}[!t]
\centering

\includegraphics[width=0.99\linewidth]{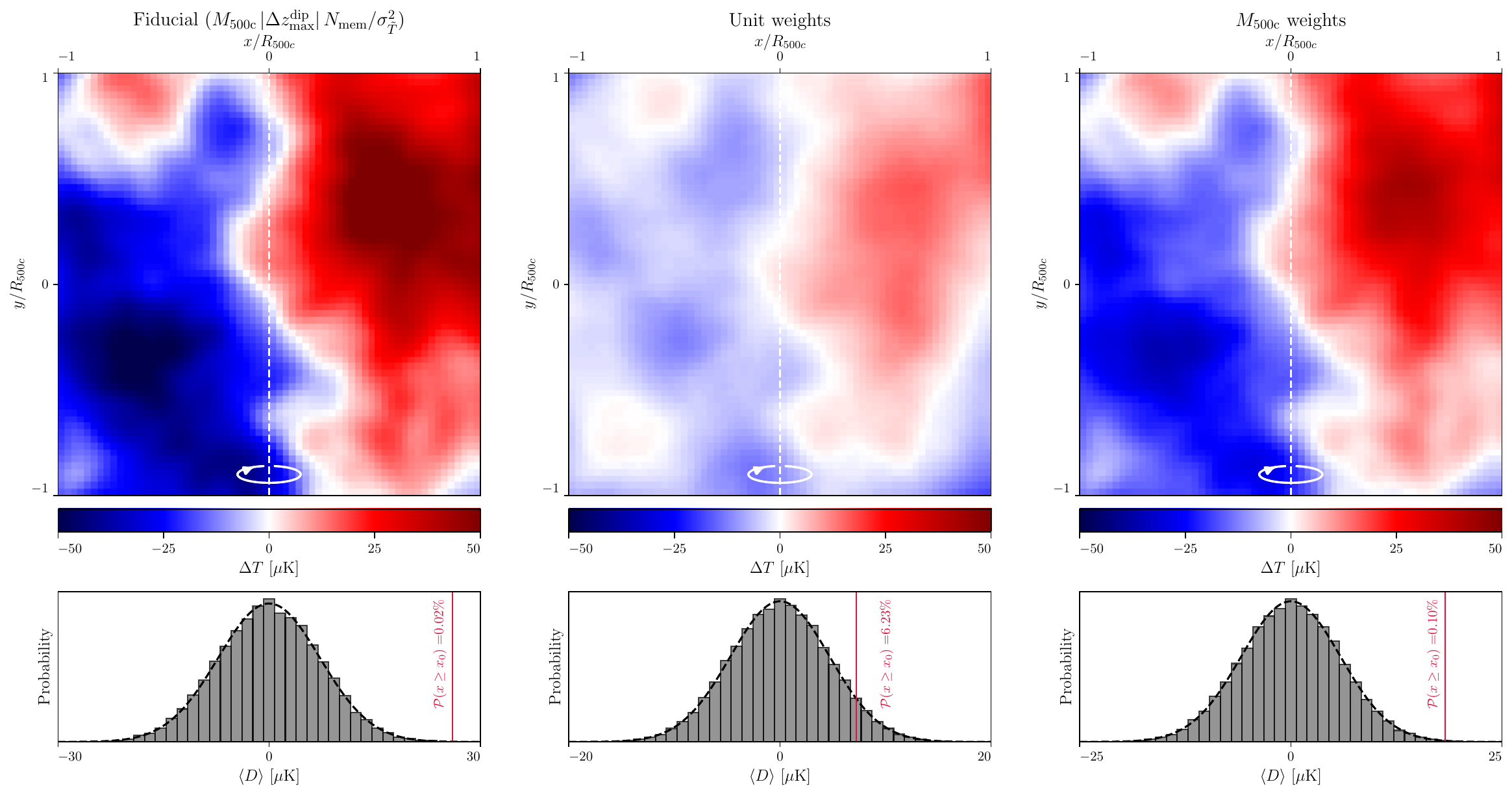}\\[0.6em]

\includegraphics[width=0.99\linewidth]{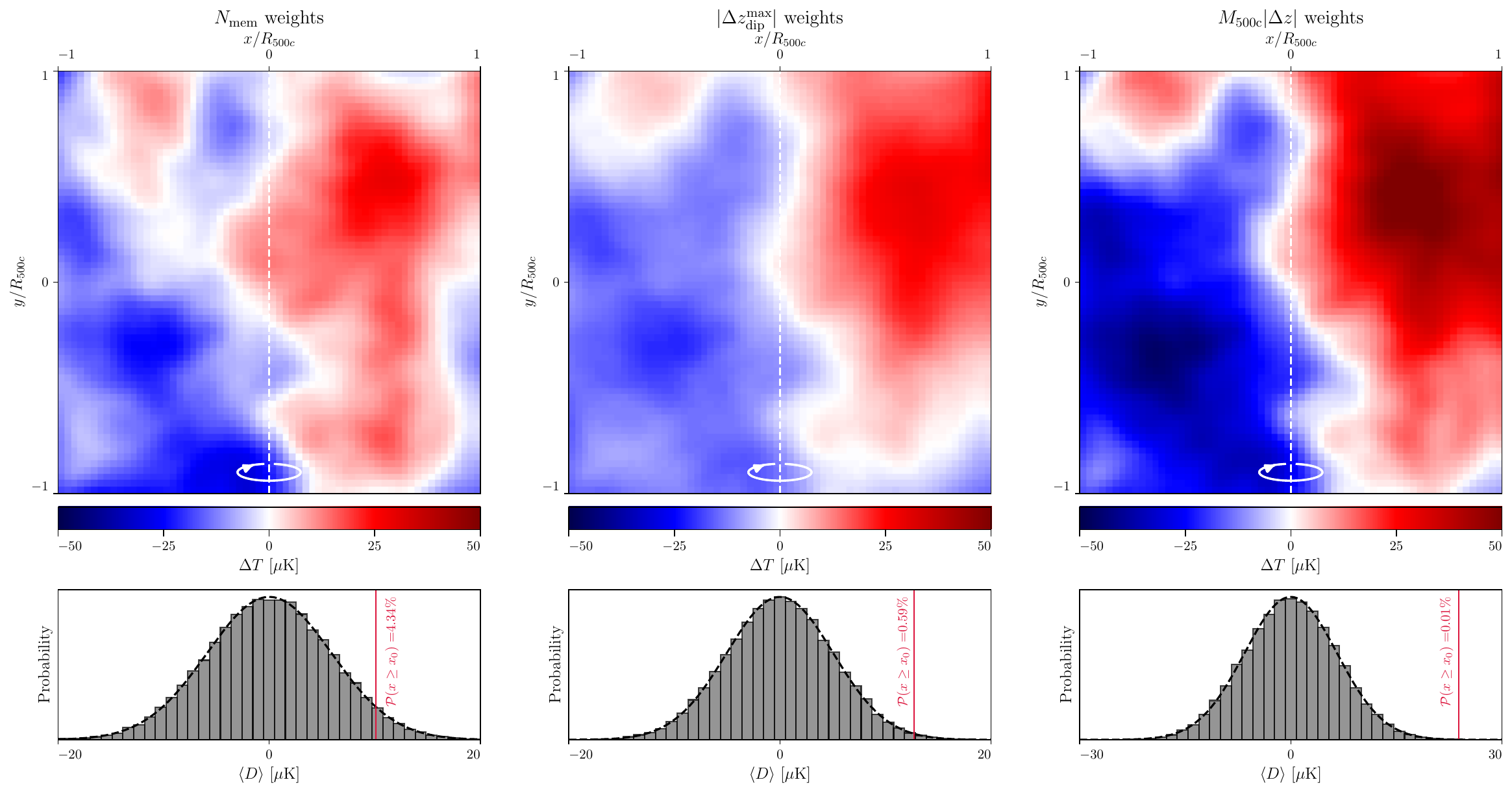}

\caption{Impact of varying weights in the oriented stack. The top row compares our fiducial analysis (left) with a stack using unit weights (middle) and a simple mass weighting (right). Mass weighting increases the dipole significance substantially. The bottom row shows a richness weighting (left), rotational velocity weighting (middle), and signal ($M\times v_{\rm rot}$) weighting (right). Incorporating an estimate of the expected signal in the weights is essential for robustly extracting the rkSZ signal.}
\label{fig:vary_weights}
\end{figure}

\clearpage
\pagebreak
\subsection{Varying weights}

\noindent In our fiducial analysis, we adopt an approximate signal-to-noise ratio weighting:
\begin{equation}\label{eq:fid_weights_SM}
    w=\frac{M_{\rm 500c}\, |\Delta z_{\rm dip}^{\rm max}| N_{\rm mem}}{\sigma^2_{\tilde{T}}(M_{\rm 500c}, z)}.
\end{equation}
Fig.~\ref{fig:vary_weights} compares several oriented stacks using different weighting procedures. Without any weighting, the dipole is insignificant $(1.5\sigma)$, which is expected given that our clusters span nearly an order of magnitude in mass. Instead, if we adopt a simple mass weighting, the significance increases to $3.1\sigma$. Conversely, using the richness as a mass proxy, i.e., weighting by $N_{\rm mem}$,\footnote{In practice, we use the geometric mean of the number of galaxies in the left and right hemispheres, $N_{\rm mem}=\sqrt{N_{\rm left}N_{\rm right}}$, since this is the quantity that is most directly related to the uncertainty on the inferred rotation axis. We have checked that our conclusions are unchanged if we use $N_{\rm mem}=N_{\rm left}+N_{\rm right}$.} leads to a negligible increase ($0.2\sigma$) in the dipole significance compared to no weighting. The effectiveness of mass weighting compared to richness weighting highlights a key advantage of cross-matching with the X-ray catalog, which provides direct cluster mass estimates. Otherwise, the weights would rely on the richness-mass relation, which can have relatively large scatter. Weighting by $|\Delta z^{\rm max}_{\rm dip}|$, which is our estimate of the cluster's rotational velocity, increases the significance of the dipole by $1\sigma$. Finally, weighting by the expected signal, $M_{\rm 500c} |\Delta z_{\rm dip}^{\rm max}|$, leads to significant $(3.7\sigma)$ evidence of a dipole, which is slightly $(0.1\sigma)$ higher than the significance of our fiducial analysis. Nevertheless, we choose to use the weights in Eq.~\eqref{eq:fid_weights_SM} in our analysis to down-weight clusters that could have noisy rotation estimates or larger contributions from the primary CMB and instrumental noise. 

\clearpage
\pagebreak

\begin{figure}[!t]
\centering

\includegraphics[width=0.99\linewidth]{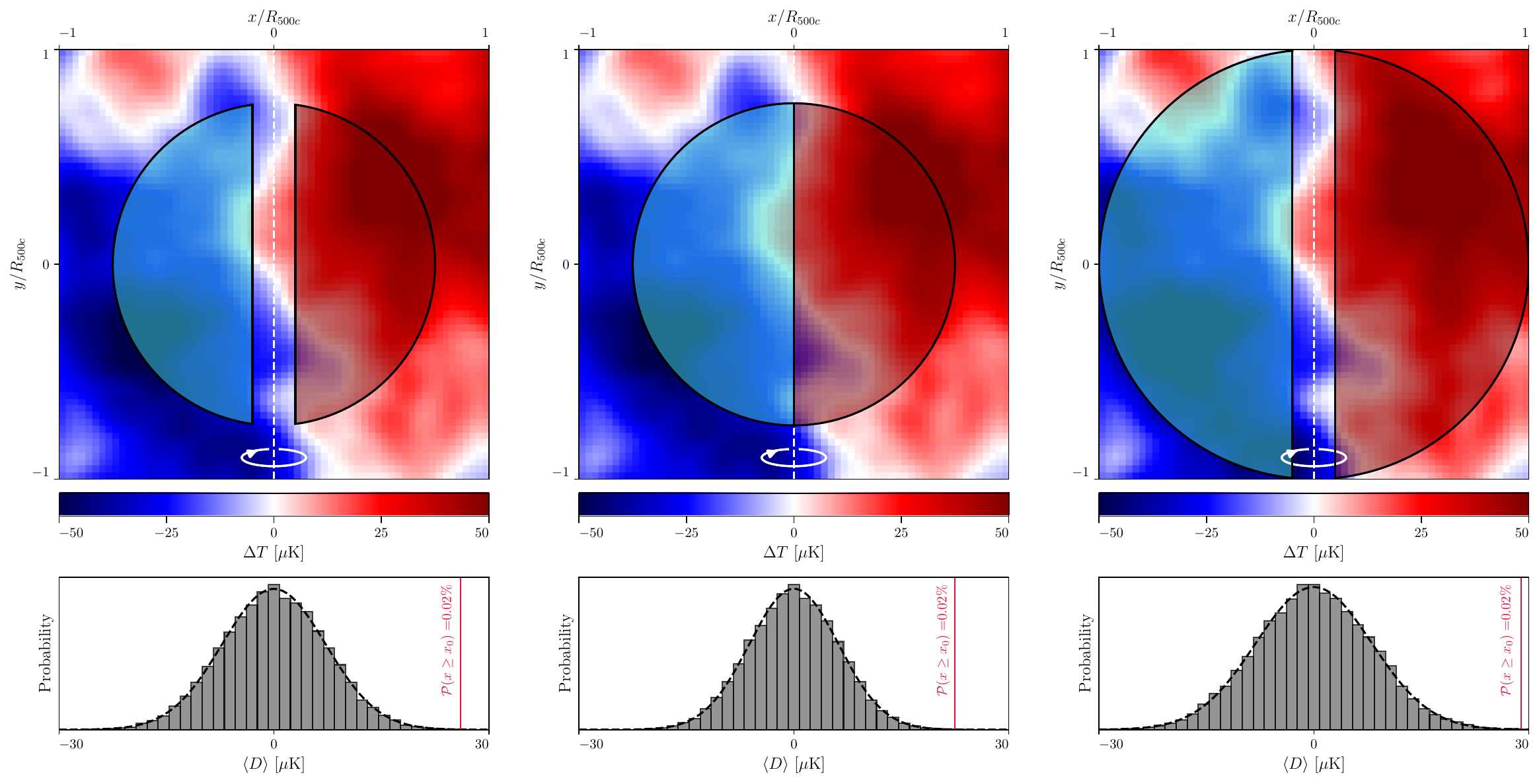}\\[0.6em]

\includegraphics[width=0.99\linewidth]{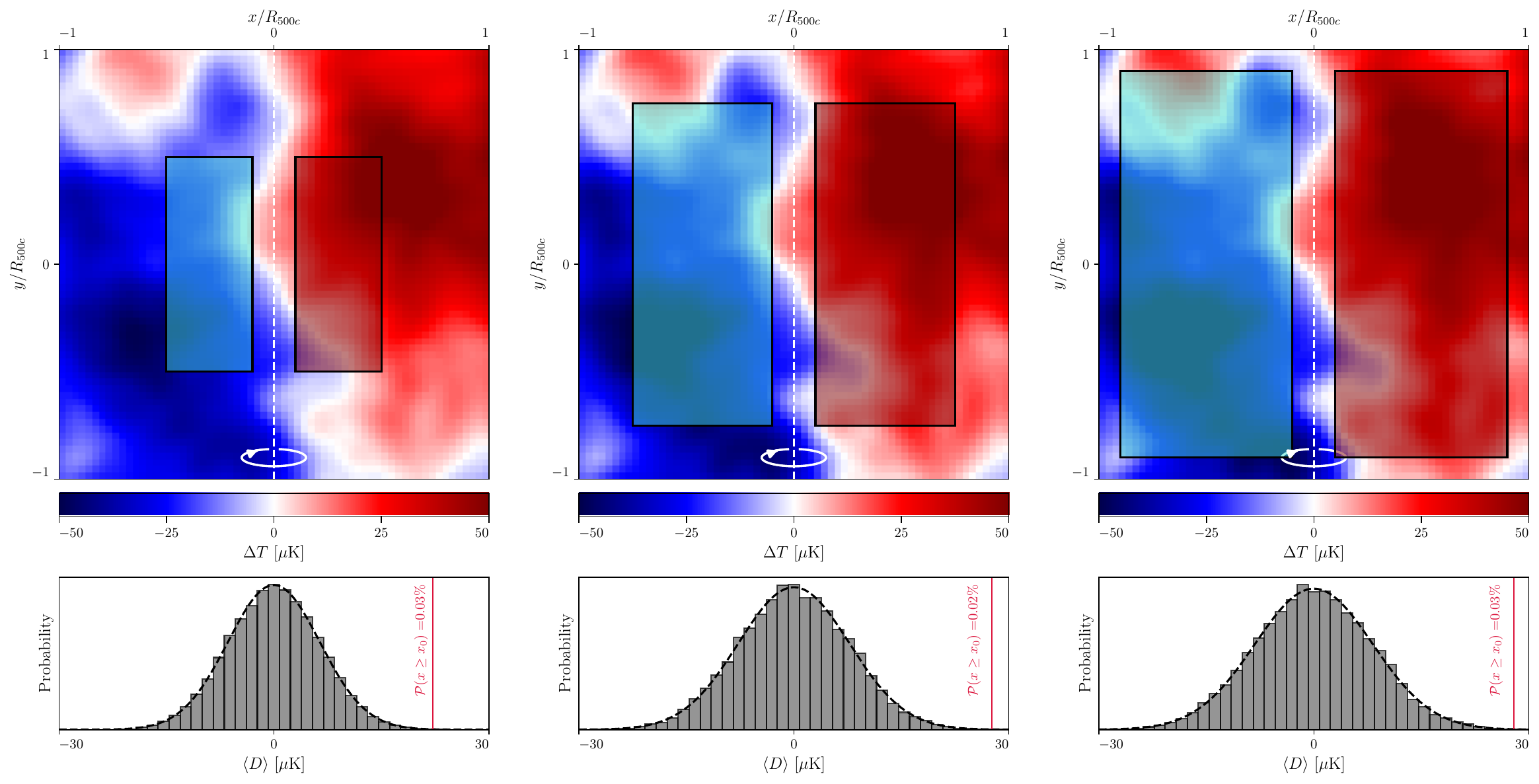}

\caption{Impact of varying the pixel-space filter size and shape on the inferred dipole amplitude, $\langle D\rangle.$ We compute $\langle D\rangle$ as the difference between the mean temperature in the right (red-shaded) and left (blue-shaded) regions of the map, relative to the vertical axis. The top-left panel shows the fiducial pixel domain used in our main analysis, which includes pixels with $r<0.75\,R_{\rm 500c}$ and $|x|\geq 0.1 R_{\rm 500c}.$ The top (bottom) row uses circular (rectangular) domains with different sizes. The inferred dipole significance is robust to variations in the pixel-space filter size and shape.}
\label{fig:vary_filters}
\end{figure}
%

\subsection{Varying pixel-space filter}

\noindent In Fig.~\ref{fig:vary_filters}, we analyze our fiducial stack using different pixel-space filters to estimate the dipole significance. For our fiducial analysis, we use pixels with $r<0.75\,R_{\rm 500c}$ and $|x|\geq 0.1\,R_{\rm 500c}$. We exclude pixels with $|x| \leq 0.1\,R_{\rm 500c}$ to minimize contributions that could arise from misidentifying the cluster centers. Including these pixels does not impact the significance of the estimated dipole, as shown in the top-middle panel. Similarly, in the top-right panel, we show that extending our analysis to include pixels out to $R_{\rm 500c}$ does not change the estimated dipole significance. However, we find that the significance decreases if we use cutouts that extend beyond $R_{\rm 500c}$, indicating that our dipole signal does not extend significantly beyond the region considered in this work. The bottom panels use rectangular pixel-space filters of varying sizes, with the bottom-left analysis closely mirroring the choices adopted in Ref.~\cite{Baxter:2019tze}. In all cases, we find significant evidence for a dipole. In conclusion, the dipolar signal presented in the \emph{Letter} is not an artifact of the pixel-space filter.

\clearpage
\pagebreak

\begin{figure}[!t]
\centering
\includegraphics[width=0.99\linewidth]{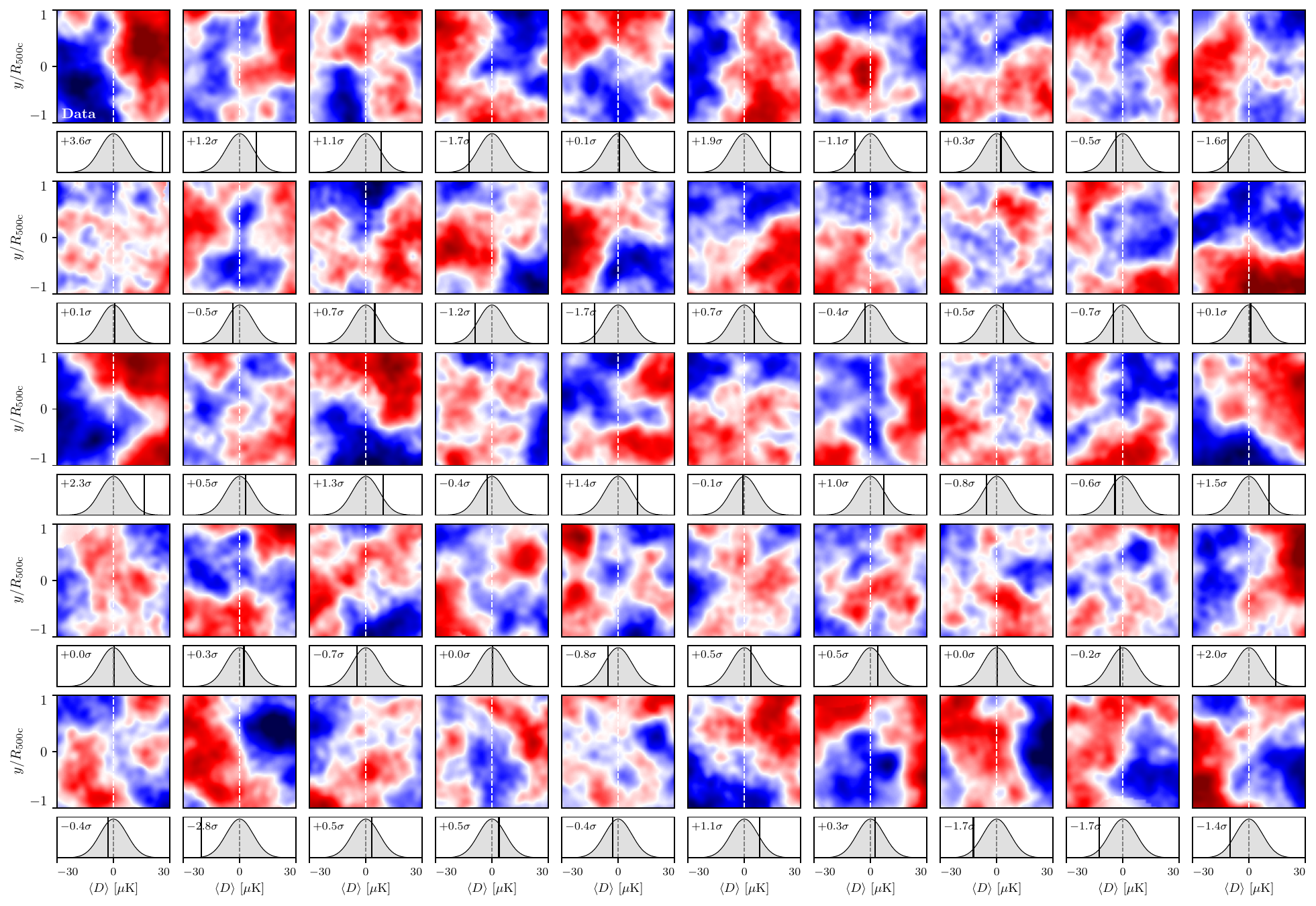}
\caption{Comparison of stack on data (top left) with 49 weighted stacks using randomly selected positions across the analysis footprint. On average, the randoms have no preferred orientation. Whereas chance alignments of fluctuations due to the primary CMB and instrumental noise can produce apparent dipoles, their amplitudes are generally much smaller than the signal presented in the \emph{Letter}, as shown by the noise distributions below each stack.}
\label{fig:random_cutouts}
\end{figure}
\begin{figure}[!t]
\centering
\includegraphics[width=0.8\linewidth]{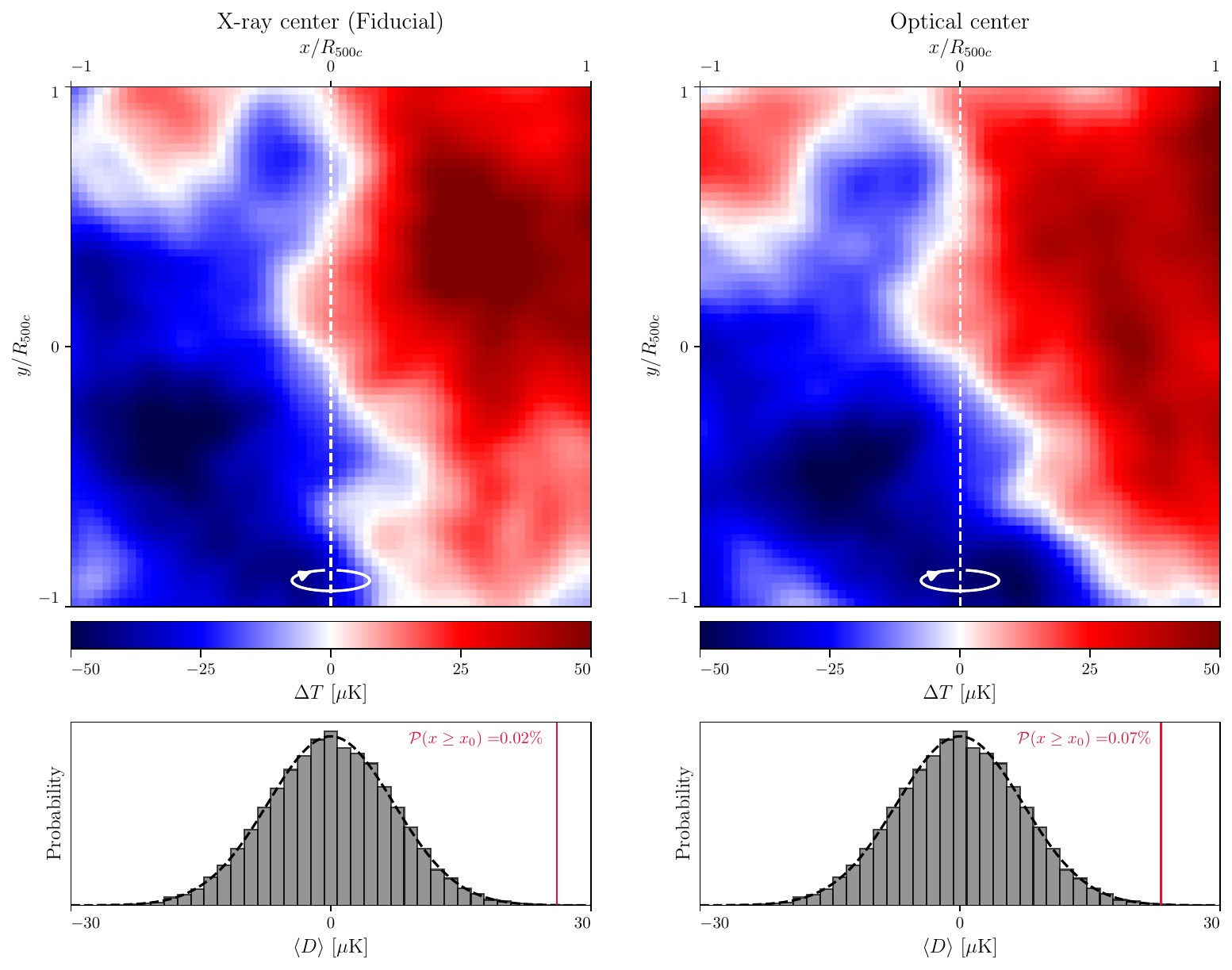}
\caption{Impact of assumed cluster center on the stacked rkSZ signal. The left panel shows our fiducial analysis, which uses the X-ray cluster center. The right panel shows the same analysis, but using the optical cluster center to both identify the rotation angle and extract cutouts. We find consistent results using the optical center, albeit with slightly lower significance ($3.2\sigma$).}
\label{fig:miscentering_stack}
\end{figure}
%

\subsection{Optical vs. X-ray center}

\noindent In this section, we assess the sensitivity of our results to the choice of cluster center. Since the kSZ signal depends on both the electron number density and the line-of-sight velocity, it can in principle be sensitive to miscentering. For example, consider a spherically symmetric cluster moving along the line of sight. If the assumed center used in the stack is offset from the true center, the resulting monopole kSZ signal from the bulk motion can project into an apparent dipole, due to the asymmetry of the electron number density with respect to the assumed center.

In our fiducial analysis, we use the X-ray center of each cluster to define the origin of our coordinate system, and hence the estimated ``rotation" axis. As shown in Fig.~\ref{fig:tSZ_stacks}, the stacked tSZ signal shows no clear evidence of miscentering. To further assess the impact of miscentering, repeat the stack using optical centers from the galaxy cluster catalog instead of the X-ray centers for our fiducial sample of 25 clusters in Fig.~\ref{fig:miscentering_stack}. Using the optical centers, we find a slightly smaller dipole with a significance of $3.2\sigma$, as opposed to $3.6\sigma$ in our fiducial analysis.

\subsection{Additional details regarding the noise model}

\noindent In our fiducial analysis, we quantify the dipole significance by repeating our stack on randomly drawn locations from the component-separated, E-mode-subtracted blackbody temperature map. Fig.~\ref{fig:random_cutouts} compares our fiducial oriented stack at the cluster locations (top left) with 49 randomly drawn stacks from our sample of 30,000 randoms. The oriented stack on cluster locations has a significantly more pronounced dipole than the randoms. Note, however, that the randoms can have apparent dipoles that are aligned with the origin of the cutouts and, in some cases, the vertical axis. These chance alignments arise because, on the angular scales used in this analysis, the primary CMB is approximately a gradient. After subtracting an overall monopole, a gradient looks like a dipole centered at the origin.

The noise model implicit in our random distribution used to assess the rkSZ significance includes the primary CMB and instrumental noise, which are the dominant contributions for the datasets and scales considered here, but it neglects any source of cluster-correlated noise.  Ref.~\cite{Baxter:2019tze} also made this approximation; nevertheless, we consider the impact of cluster-correlated noise in more detail here.  In principle, one could estimate cluster-correlated noise using a (non-oriented) stack on a sample of randomly selected galaxy clusters that accurately reproduces the characteristics of the cluster sample used in the signal stack (e.g., mass distribution, redshift distribution, etc.). In practice, this is infeasible given the small number of massive, low-redshift ($z<0.1$) clusters in the sky.  Alternatively, one could use simulated covariances that include contributions from, e.g., the CMB, kSZ, tSZ, dust emission, radio source emission, lensing, the moving lens effect, and instrumental noise, but constructing such simulations is beyond the scope of this work, and as noted below the contributions from noise sources other than the primary CMB and instrumental noise are expected to be negligible here.

To roughly estimate the impact of cluster-correlated noise, we repeat our stack 30,000 times on the true cluster positions using random rotation angles. We stress that this procedure does not constitute a formal assessment of the detection significance because the resulting noise distribution is amplified by the signal, i.e., the signal and noise are highly correlated here, as some of the signal will frequently leak into the random stacks in this procedure.  Moreover, an optimal analysis based on this method would include different weights than Eq.~\eqref{eq:fid_weights_SM}, which were designed to minimize contributions from the primary CMB and instrumental noise. Thus, this approach will underestimate the true detection significance. Using the randomized rotation angles, we still find significant ($2.9\sigma$) evidence of a dipole, indicating that contributions from cluster-correlated foregrounds are subdominant compared to the primary CMB and instrumental noise. On the other hand, future rkSZ analyses using higher-resolution CMB data will require a dedicated treatment of cluster-correlated noise, as these experiments will probe scales with significantly smaller contributions from the primary CMB and instrumental noise than the present analysis.

\clearpage
\pagebreak

\section{Profile theory model and measurement details}

%
\subsection{Derivation of profile model}

\noindent In this section, we derive the theoretical model used to fit the azimuthally-averaged rotational kSZ profile within each hemisphere, as presented in the \emph{Letter}. The rotational kSZ temperature fluctuation produced by a cluster of mass $M_{\rm 500c}$ at redshift $z$ at a projected radius $R$ and azimuthal angle $\theta$ (measured clockwise from the rotation direction, taken to be the $y$-axis) is
\begin{equation} 
T_{\rm rkSZ}(R,\theta)=\frac{\sigma_T}{c}\,T_{\rm CMB}\int_{-\infty}^{\infty} dl\;n_e\left(\sqrt{R^2+l^2}\,|\,M_{\rm 500c},z\right) v_{\rm rot,\parallel}(R,\theta),
\end{equation}
where $l$ is the line-of-sight distance, $n_e(r\,|\,M_{\rm 500c},z)$ is the 3D electron number density profile, and $v_{\rm rot,\parallel}$ is the line-of-sight component of the rotational velocity. Assuming a solid-body rotation model with angular velocity $\omega$ and $v_{\rm rot,\parallel}(R,\theta) = R\omega\sin\theta$ and a spherically-symmetric electron number density profile, the rotational kSZ signal is 
\begin{equation}
     T_{\rm rkSZ}(R,\theta)=\frac{2\,T_{\rm CMB}\,\sigma_T\,\omega}{c}\,R\,\sin\theta\int\limits_0^{\sqrt{r_{\rm max}^2-R^2}}\,dl\,n_e(\sqrt{R^2+l^2}\, |\,M_{\rm 500c},\,z)\equiv \frac{2\,T_{\rm CMB}\,\sigma_T\,\omega}{c}\,R\,\sin\theta\,F_e(R),
\end{equation}
where $r_{\rm max}$ is the maximum 3D radius to which we integrate the profile (i.e., this is the halo boundary), and we have defined the projected electron number density profile, $F_e(R)\equiv\int_0^{\sqrt{r_{\rm max}-R^2}}\,dl\,n_e(\sqrt{R^2+l^2}\, |\,M_{\rm 500c},\,z)$, for convenience.

Assuming a Gaussian beam with standard deviation $\sigma_b$ and using the flat-sky approximation, the beam-convolved rotational kSZ signal is 
\begin{align}
\hat{T}_{\rm rkSZ}(R,\theta)&=\frac{1}{2\pi \sigma_b^2}\int_{0}^{R_{\rm max}}dR'\, R'\int\limits_{0}^{2\pi}d\theta' \,T_{\rm rkSZ}(R', \theta')\exp\left[-\frac{R'^2+R^2-2R'R\cos(\theta-\theta')}{2\sigma_b^2} \right] ,\\
&=\frac{2\sigma_T\,T_{\rm CMB}\,\omega}{c}\frac{1}{2\pi\sigma_b^2}\int_{0}^{R_{\rm max}} dR'\, R'^2 F_e(R')
\int_{0}^{2\pi} d\theta'\;
\sin\theta'\,
\exp\!\left[
-\frac{R'^2+R^2-2R'R\cos(\theta-\theta')}
{2\sigma_b^2}
\right],
\end{align}
where $R_{\rm max}$ is the maximum 2D projected radius to which we integrate. The angular integral can be performed analytically
\begin{equation}
\int\limits_{0}^{2\pi}d\theta'\; \sin\theta'\;\left(\exp\left[-\frac{R'^2+R^2-2R'R\cos(\theta-\theta')}{2\sigma_b^2} \right] \right)=2 \pi \, \sin\theta \, I_1\Big(\frac{R' R}{\sigma_b^2}\Big)e^{-\frac{R'^2+R^2}{2\sigma_b^2}},
\end{equation}
where $I_1$ is the first modified Bessel function of the first kind. Therefore, the beam-convolved rotational kSZ signal for a solid-body rotation model is
\begin{equation}
\hat{T}_{\rm rkSZ}(R, \theta)=\left(\frac{2\sigma_T\,T_{\rm CMB}}{c} \right)\frac{\omega \sin\theta}{\sigma_b^2}\int\limits_{0}^{R_{\rm max}}dR'\, R'^2 F_e(R') I_1\Big(\frac{R'R}{\sigma_b^2}\Big) \,e^{-\frac{R'^2+R^2}{2\sigma_b^2}}
\end{equation}
Finally, taking the azimuthal average over each hemisphere, we find
\begin{equation}\label{eq:rkSZ_profile}
\hat{T}_{\rm rkSZ}(R)=\pm \left(\frac{4\sigma_T\,T_{\rm CMB}}{\pi c} \right)\frac{\omega}{\sigma_b^2}\int\limits_{0}^{R_{\rm max}}dR'\, R'^2 F_e(R') \,I_1\Big(\frac{R'R}{\sigma_b^2}\Big) \,e^{-\frac{R'^2+R^2}{2\sigma_b^2}},
\end{equation}
where the $\pm$ sign depends on which hemisphere one integrates over.

We compute the theoretical profile in Fig.~\ref{fig:rkSZ_profile} of the main text numerically for each cluster in our sample using Eq.~\eqref{eq:rkSZ_profile}, assuming $r_{\rm max}=R_{\rm max}=5\,R_{\rm 500c}$. 

We model the electron distribution within each cluster using the hydrodynamical-simulation-based ``AGN Feedback" profile from~\citet{Battaglia:2016xbi}. This profile models the gas density as a generalized NFW (gNFW) profile:\footnote{Assuming the plasma is fully ionized, the electron number density is related to the total gas density by $n_e(r)=\frac{1+X_H}{2 m_u}\rho_{\rm gas}(r),$ where $m_u$ is the atomic mass unit and $X_H\approx 0.76$ is the hydrogen mass fraction.}
\begin{equation}\label{eq:rho_gas_battaglia}
 \rho_{\rm gas}(r)=f_b\, \rho_{\rm crit}(z)\, \rho_0
\,  \left(\frac{x}{x_c}\right) ^\gamma\left[1+\left(\frac{x}{x_c}\right)^\alpha \right]^{-\frac{\beta+\gamma}{\alpha}}
\end{equation}
where $f_b\equiv \Omega_b/\Omega_m$ is the baryon fraction, $\rho_{\rm crit}$ is the critical density, $x\equiv r/R_{200c}$, and we fix $x_c=0.5$ and $\gamma=-0.2$. In Eq.~\eqref{eq:rho_gas_battaglia}, $\rho_0$, $\alpha$, and $\beta$ are mass- and redshift-dependent quantities parameterized by 
\begin{equation}
    X(M_{200c},z)=X_0\left( \frac{M_{200c}}{10^{14}~M_\odot}\right)^{X_M}(1+z)^{X_z} \text{ for } X\in \{\rho_0,\,\alpha,\,\beta\},
\end{equation}
where the best-fit values of $X_0,$ $X_M$, and $X_z$ are given in Table 2 of~\cite{Bolliet:2022pze}. To compute the profile for the stacked signal, we compute the profile for each cluster, rescale the projected radial coordinate to be in units of $R/R_{\rm 500c}$, and take the weighted mean profile of this sample. We assume each cluster is rotating at $1500~{\rm km/s}$ at a radius of $0.2 r_{\rm 500c}$ so that our theory template has a peak amplitude of $\sim 25~{\mu {\rm K}}$ to match the predictions with those from simulations~\cite{Altamura:2023hoe}. We convert from $M_{\rm 500c}$ and $R_{\rm 500c}$ to $M_{\rm 200c}$ and $R_{\rm 200c}$ using the concentration-mass relation from Ref.~\cite{Bhattacharya:2011vr}.

\begin{figure}[!t]
\centering
\includegraphics[width=0.8\linewidth]{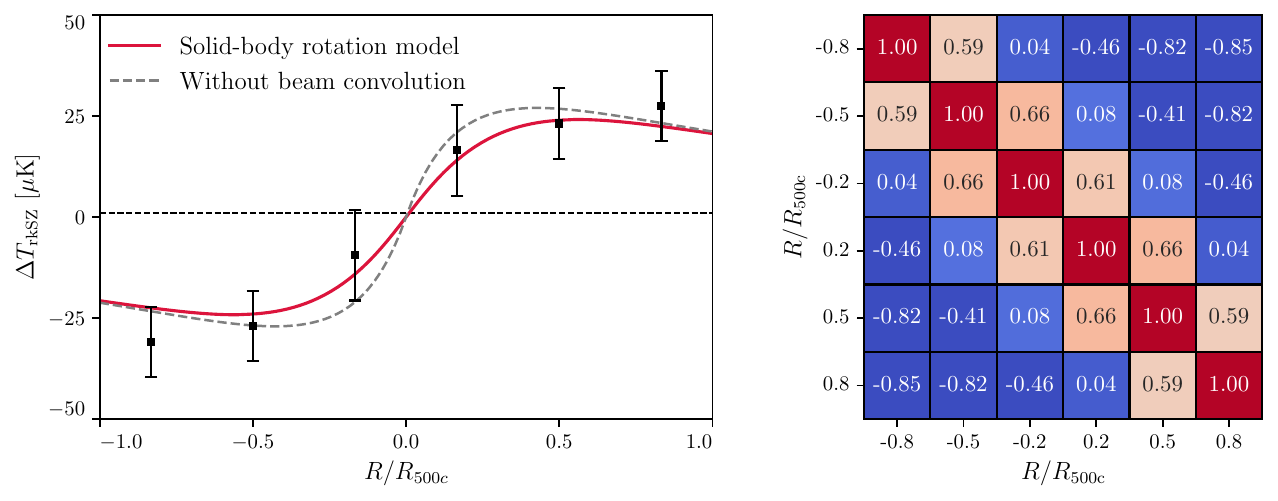}
\caption{\emph{Left}: impact of beam convolution on the theoretical rkSZ profile. \emph{Right}: correlation matrix for the measured rkSZ profile. The adjacent bins have a large positive correlation because of the projection from Fourier to real space. Bins with larger separations are anti-correlated because the primary CMB is approximately a gradient on these angular scales.}
\label{fig:prof_corr_mat}
\end{figure}

The left panel of Fig.~\ref{fig:prof_corr_mat} shows our measurement of the rkSZ profile compared with theoretical predictions, both with and without beam convolution. As expected, the dominant effect of the beam is to flatten the profile.

\subsection{Additional details on profile measurement and fitting}

\noindent We fit the measured profile assuming a Gaussian likelihood and rescale the theoretical template (Eq.~\eqref{eq:rkSZ_profile}) by a dimensionless amplitude, $A_{\rm rkSZ},$ 
\begin{equation}
   \Delta T^{\rm obs}_{\rm rkSZ}(R)=A_{\rm rkSZ}\times \hat{T}_{\rm rkSZ}(R),
\end{equation}
with a uniform prior on the amplitude, $-5 \leq A_{\rm kSZ} \leq 5$. The marginalized posterior is $A_{\rm rkSZ}=1.05\pm 0.32$ at the 68\% two-tailed confidence limit, corresponding to a 3.3$\sigma$ preference for a non-zero amplitude. We emphasize that this is not a constraint on the physical rkSZ amplitude, because we fix the rotational velocity to produce a dipole of order $25~\mu{\rm K}$ and do not directly account for signal loss due to projection and angle misalignment effects. Instead, this profile analysis offers an alternative way to quantify the dipole significance and provides a qualitative comparison of the measured profile’s shape and amplitude to those predicted by simulations. The solid-body rotation model fits the azimuthally-averaged profile well, with a probability-to-exceed (PTE) of 0.51.

The right panel of Fig.~\ref{fig:prof_corr_mat} shows the correlation matrix of the azimuthally-averaged profile in each hemisphere, estimated from 30,000 random stacks. The data points are highly correlated. On small scales, neighboring bins are correlated because of the projection from harmonic- to real-space. On large scales, the bins are anti-correlated because, on these scales, the primary CMB is approximately a gradient.

\end{document}